\documentclass[journal,cspaper,compsoc,final]{IEEEtran}

\usepackage{ifpdf}
\ifCLASSOPTIONcompsoc
  \usepackage[nocompress]{cite}
\else  
  \usepackage{cite}
\fi

\usepackage[cmex10]{amsmath}
\usepackage[notheorems]{rr-math}

\newcommand{\eat}[1]{}

\usepackage{multirow}
\usepackage{comment}
\usepackage{rotating}
\usepackage{afterpage}

\usepackage{graphicx}
\usepackage{epsfig}
\usepackage{graphics}
\usepackage{amssymb}
\usepackage{amsmath}
\usepackage{rotate}
\usepackage{color}
\usepackage{subfigure}
\usepackage{epstopdf}

\usepackage{tabulary}

\definecolor{gray}{RGB}{150,150,150}
  \definecolor{snow}{rgb}{1.000000,0.980392,0.980392}
  \definecolor{ghost white}{rgb}{0.972549,0.972549,1.000000}
  \definecolor{GhostWhite}{rgb}{0.972549,0.972549,1.000000}
  \definecolor{white smoke}{rgb}{0.960784,0.960784,0.960784}
  \definecolor{WhiteSmoke}{rgb}{0.960784,0.960784,0.960784}
  \definecolor{gainsboro}{rgb}{0.862745,0.862745,0.862745}
  \definecolor{floral white}{rgb}{1.000000,0.980392,0.941176}
  \definecolor{FloralWhite}{rgb}{1.000000,0.980392,0.941176}
  \definecolor{old lace}{rgb}{0.992157,0.960784,0.901961}
  \definecolor{OldLace}{rgb}{0.992157,0.960784,0.901961}
  \definecolor{linen}{rgb}{0.980392,0.941176,0.901961}
  \definecolor{antique white}{rgb}{0.980392,0.921569,0.843137}
  \definecolor{AntiqueWhite}{rgb}{0.980392,0.921569,0.843137}
  \definecolor{papaya whip}{rgb}{1.000000,0.937255,0.835294}
  \definecolor{PapayaWhip}{rgb}{1.000000,0.937255,0.835294}
  \definecolor{blanched almond}{rgb}{1.000000,0.921569,0.803922}
  \definecolor{BlanchedAlmond}{rgb}{1.000000,0.921569,0.803922}
  \definecolor{bisque}{rgb}{1.000000,0.894118,0.768627}
  \definecolor{peach puff}{rgb}{1.000000,0.854902,0.725490}
  \definecolor{PeachPuff}{rgb}{1.000000,0.854902,0.725490}
  \definecolor{navajo white}{rgb}{1.000000,0.870588,0.678431}
  \definecolor{NavajoWhite}{rgb}{1.000000,0.870588,0.678431}
  \definecolor{moccasin}{rgb}{1.000000,0.894118,0.709804}
  \definecolor{cornsilk}{rgb}{1.000000,0.972549,0.862745}
  \definecolor{ivory}{rgb}{1.000000,1.000000,0.941176}
  \definecolor{lemon chiffon}{rgb}{1.000000,0.980392,0.803922}
  \definecolor{LemonChiffon}{rgb}{1.000000,0.980392,0.803922}
  \definecolor{seashell}{rgb}{1.000000,0.960784,0.933333}
  \definecolor{honeydew}{rgb}{0.941176,1.000000,0.941176}
  \definecolor{mint cream}{rgb}{0.960784,1.000000,0.980392}
  \definecolor{MintCream}{rgb}{0.960784,1.000000,0.980392}
  \definecolor{azure}{rgb}{0.941176,1.000000,1.000000}
  \definecolor{alice blue}{rgb}{0.941176,0.972549,1.000000}
  \definecolor{AliceBlue}{rgb}{0.941176,0.972549,1.000000}
  \definecolor{lavender}{rgb}{0.901961,0.901961,0.980392}
  \definecolor{lavender blush}{rgb}{1.000000,0.941176,0.960784}
  \definecolor{LavenderBlush}{rgb}{1.000000,0.941176,0.960784}
  \definecolor{misty rose}{rgb}{1.000000,0.894118,0.882353}
  \definecolor{MistyRose}{rgb}{1.000000,0.894118,0.882353}
  \definecolor{white}{rgb}{1.000000,1.000000,1.000000}
  \definecolor{black}{rgb}{0.000000,0.000000,0.000000}
  \definecolor{dark slate gray}{rgb}{0.184314,0.309804,0.309804}
  \definecolor{DarkSlateGray}{rgb}{0.184314,0.309804,0.309804}
  \definecolor{dark slate grey}{rgb}{0.184314,0.309804,0.309804}
  \definecolor{DarkSlateGrey}{rgb}{0.184314,0.309804,0.309804}
  \definecolor{dim gray}{rgb}{0.411765,0.411765,0.411765}
  \definecolor{DimGray}{rgb}{0.411765,0.411765,0.411765}
  \definecolor{dim grey}{rgb}{0.411765,0.411765,0.411765}
  \definecolor{DimGrey}{rgb}{0.411765,0.411765,0.411765}
  \definecolor{slate gray}{rgb}{0.439216,0.501961,0.564706}
  \definecolor{SlateGray}{rgb}{0.439216,0.501961,0.564706}
  \definecolor{slate grey}{rgb}{0.439216,0.501961,0.564706}
  \definecolor{SlateGrey}{rgb}{0.439216,0.501961,0.564706}
  \definecolor{light slate gray}{rgb}{0.466667,0.533333,0.600000}
  \definecolor{LightSlateGray}{rgb}{0.466667,0.533333,0.600000}
  \definecolor{light slate grey}{rgb}{0.466667,0.533333,0.600000}
  \definecolor{LightSlateGrey}{rgb}{0.466667,0.533333,0.600000}
  \definecolor{gray}{rgb}{0.745098,0.745098,0.745098}
  \definecolor{grey}{rgb}{0.745098,0.745098,0.745098}
  \definecolor{light grey}{rgb}{0.827451,0.827451,0.827451}
  \definecolor{LightGrey}{rgb}{0.827451,0.827451,0.827451}
  \definecolor{light gray}{rgb}{0.827451,0.827451,0.827451}
  \definecolor{LightGray}{rgb}{0.827451,0.827451,0.827451}
  \definecolor{midnight blue}{rgb}{0.098039,0.098039,0.439216}
  \definecolor{MidnightBlue}{rgb}{0.098039,0.098039,0.439216}
  \definecolor{navy}{rgb}{0.000000,0.000000,0.501961}
  \definecolor{navy blue}{rgb}{0.000000,0.000000,0.501961}
  \definecolor{NavyBlue}{rgb}{0.000000,0.000000,0.501961}
  \definecolor{cornflower blue}{rgb}{0.392157,0.584314,0.929412}
  \definecolor{CornflowerBlue}{rgb}{0.392157,0.584314,0.929412}
  \definecolor{dark slate blue}{rgb}{0.282353,0.239216,0.545098}
  \definecolor{DarkSlateBlue}{rgb}{0.282353,0.239216,0.545098}
  \definecolor{slate blue}{rgb}{0.415686,0.352941,0.803922}
  \definecolor{SlateBlue}{rgb}{0.415686,0.352941,0.803922}
  \definecolor{medium slate blue}{rgb}{0.482353,0.407843,0.933333}
  \definecolor{MediumSlateBlue}{rgb}{0.482353,0.407843,0.933333}
  \definecolor{light slate blue}{rgb}{0.517647,0.439216,1.000000}
  \definecolor{LightSlateBlue}{rgb}{0.517647,0.439216,1.000000}
  \definecolor{medium blue}{rgb}{0.000000,0.000000,0.803922}
  \definecolor{MediumBlue}{rgb}{0.000000,0.000000,0.803922}
  \definecolor{royal blue}{rgb}{0.254902,0.411765,0.882353}
  \definecolor{RoyalBlue}{rgb}{0.254902,0.411765,0.882353}
  \definecolor{blue}{rgb}{0.000000,0.000000,1.000000}
  \definecolor{dodger blue}{rgb}{0.117647,0.564706,1.000000}
  \definecolor{DodgerBlue}{rgb}{0.117647,0.564706,1.000000}
  \definecolor{deep sky blue}{rgb}{0.000000,0.749020,1.000000}
  \definecolor{DeepSkyBlue}{rgb}{0.000000,0.749020,1.000000}
  \definecolor{sky blue}{rgb}{0.529412,0.807843,0.921569}
  \definecolor{SkyBlue}{rgb}{0.529412,0.807843,0.921569}
  \definecolor{light sky blue}{rgb}{0.529412,0.807843,0.980392}
  \definecolor{LightSkyBlue}{rgb}{0.529412,0.807843,0.980392}
  \definecolor{steel blue}{rgb}{0.274510,0.509804,0.705882}
  \definecolor{SteelBlue}{rgb}{0.274510,0.509804,0.705882}
  \definecolor{light steel blue}{rgb}{0.690196,0.768627,0.870588}
  \definecolor{LightSteelBlue}{rgb}{0.690196,0.768627,0.870588}
  \definecolor{light blue}{rgb}{0.678431,0.847059,0.901961}
  \definecolor{LightBlue}{rgb}{0.678431,0.847059,0.901961}
  \definecolor{powder blue}{rgb}{0.690196,0.878431,0.901961}
  \definecolor{PowderBlue}{rgb}{0.690196,0.878431,0.901961}
  \definecolor{pale turquoise}{rgb}{0.686275,0.933333,0.933333}
  \definecolor{PaleTurquoise}{rgb}{0.686275,0.933333,0.933333}
  \definecolor{dark turquoise}{rgb}{0.000000,0.807843,0.819608}
  \definecolor{DarkTurquoise}{rgb}{0.000000,0.807843,0.819608}
  \definecolor{medium turquoise}{rgb}{0.282353,0.819608,0.800000}
  \definecolor{MediumTurquoise}{rgb}{0.282353,0.819608,0.800000}
  \definecolor{turquoise}{rgb}{0.250980,0.878431,0.815686}
  \definecolor{cyan}{rgb}{0.000000,1.000000,1.000000}
  \definecolor{light cyan}{rgb}{0.878431,1.000000,1.000000}
  \definecolor{LightCyan}{rgb}{0.878431,1.000000,1.000000}
  \definecolor{cadet blue}{rgb}{0.372549,0.619608,0.627451}
  \definecolor{CadetBlue}{rgb}{0.372549,0.619608,0.627451}
  \definecolor{medium aquamarine}{rgb}{0.400000,0.803922,0.666667}
  \definecolor{MediumAquamarine}{rgb}{0.400000,0.803922,0.666667}
  \definecolor{aquamarine}{rgb}{0.498039,1.000000,0.831373}
  \definecolor{dark green}{rgb}{0.000000,0.392157,0.000000}
  \definecolor{DarkGreen}{rgb}{0.000000,0.392157,0.000000}
  \definecolor{dark olive green}{rgb}{0.333333,0.419608,0.184314}
  \definecolor{DarkOliveGreen}{rgb}{0.333333,0.419608,0.184314}
  \definecolor{dark sea green}{rgb}{0.560784,0.737255,0.560784}
  \definecolor{DarkSeaGreen}{rgb}{0.560784,0.737255,0.560784}
  \definecolor{sea green}{rgb}{0.180392,0.545098,0.341176}
  \definecolor{SeaGreen}{rgb}{0.180392,0.545098,0.341176}
  \definecolor{medium sea green}{rgb}{0.235294,0.701961,0.443137}
  \definecolor{MediumSeaGreen}{rgb}{0.235294,0.701961,0.443137}
  \definecolor{light sea green}{rgb}{0.125490,0.698039,0.666667}
  \definecolor{LightSeaGreen}{rgb}{0.125490,0.698039,0.666667}
  \definecolor{pale green}{rgb}{0.596078,0.984314,0.596078}
  \definecolor{PaleGreen}{rgb}{0.596078,0.984314,0.596078}
  \definecolor{spring green}{rgb}{0.000000,1.000000,0.498039}
  \definecolor{SpringGreen}{rgb}{0.000000,1.000000,0.498039}
  \definecolor{lawn green}{rgb}{0.486275,0.988235,0.000000}
  \definecolor{LawnGreen}{rgb}{0.486275,0.988235,0.000000}
  \definecolor{green}{rgb}{0.000000,1.000000,0.000000}
  \definecolor{chartreuse}{rgb}{0.498039,1.000000,0.000000}
  \definecolor{medium spring green}{rgb}{0.000000,0.980392,0.603922}
  \definecolor{MediumSpringGreen}{rgb}{0.000000,0.980392,0.603922}
  \definecolor{green yellow}{rgb}{0.678431,1.000000,0.184314}
  \definecolor{GreenYellow}{rgb}{0.678431,1.000000,0.184314}
  \definecolor{lime green}{rgb}{0.196078,0.803922,0.196078}
  \definecolor{LimeGreen}{rgb}{0.196078,0.803922,0.196078}
  \definecolor{yellow green}{rgb}{0.603922,0.803922,0.196078}
  \definecolor{YellowGreen}{rgb}{0.603922,0.803922,0.196078}
  \definecolor{forest green}{rgb}{0.133333,0.545098,0.133333}
  \definecolor{ForestGreen}{rgb}{0.133333,0.545098,0.133333}
  \definecolor{olive drab}{rgb}{0.419608,0.556863,0.137255}
  \definecolor{OliveDrab}{rgb}{0.419608,0.556863,0.137255}
  \definecolor{dark khaki}{rgb}{0.741176,0.717647,0.419608}
  \definecolor{DarkKhaki}{rgb}{0.741176,0.717647,0.419608}
  \definecolor{khaki}{rgb}{0.941176,0.901961,0.549020}
  \definecolor{pale goldenrod}{rgb}{0.933333,0.909804,0.666667}
  \definecolor{PaleGoldenrod}{rgb}{0.933333,0.909804,0.666667}
  \definecolor{light goldenrod yellow}{rgb}{0.980392,0.980392,0.823529}
  \definecolor{LightGoldenrodYellow}{rgb}{0.980392,0.980392,0.823529}
  \definecolor{light yellow}{rgb}{1.000000,1.000000,0.878431}
  \definecolor{LightYellow}{rgb}{1.000000,1.000000,0.878431}
  \definecolor{yellow}{rgb}{1.000000,1.000000,0.000000}
  \definecolor{gold}{rgb}{1.000000,0.843137,0.000000}
  \definecolor{light goldenrod}{rgb}{0.933333,0.866667,0.509804}
  \definecolor{LightGoldenrod}{rgb}{0.933333,0.866667,0.509804}
  \definecolor{goldenrod}{rgb}{0.854902,0.647059,0.125490}
  \definecolor{dark goldenrod}{rgb}{0.721569,0.525490,0.043137}
  \definecolor{DarkGoldenrod}{rgb}{0.721569,0.525490,0.043137}
  \definecolor{rosy brown}{rgb}{0.737255,0.560784,0.560784}
  \definecolor{RosyBrown}{rgb}{0.737255,0.560784,0.560784}
  \definecolor{indian red}{rgb}{0.803922,0.360784,0.360784}
  \definecolor{IndianRed}{rgb}{0.803922,0.360784,0.360784}
  \definecolor{saddle brown}{rgb}{0.545098,0.270588,0.074510}
  \definecolor{SaddleBrown}{rgb}{0.545098,0.270588,0.074510}
  \definecolor{sienna}{rgb}{0.627451,0.321569,0.176471}
  \definecolor{peru}{rgb}{0.803922,0.521569,0.247059}
  \definecolor{burlywood}{rgb}{0.870588,0.721569,0.529412}
  \definecolor{beige}{rgb}{0.960784,0.960784,0.862745}
  \definecolor{wheat}{rgb}{0.960784,0.870588,0.701961}
  \definecolor{sandy brown}{rgb}{0.956863,0.643137,0.376471}
  \definecolor{SandyBrown}{rgb}{0.956863,0.643137,0.376471}
  \definecolor{tan}{rgb}{0.823529,0.705882,0.549020}
  \definecolor{chocolate}{rgb}{0.823529,0.411765,0.117647}
  \definecolor{firebrick}{rgb}{0.698039,0.133333,0.133333}
  \definecolor{brown}{rgb}{0.647059,0.164706,0.164706}
  \definecolor{dark salmon}{rgb}{0.913725,0.588235,0.478431}
  \definecolor{DarkSalmon}{rgb}{0.913725,0.588235,0.478431}
  \definecolor{salmon}{rgb}{0.980392,0.501961,0.447059}
  \definecolor{light salmon}{rgb}{1.000000,0.627451,0.478431}
  \definecolor{LightSalmon}{rgb}{1.000000,0.627451,0.478431}
  \definecolor{orange}{rgb}{1.000000,0.647059,0.000000}
  \definecolor{dark orange}{rgb}{1.000000,0.549020,0.000000}
  \definecolor{DarkOrange}{rgb}{1.000000,0.549020,0.000000}
  \definecolor{coral}{rgb}{1.000000,0.498039,0.313726}
  \definecolor{light coral}{rgb}{0.941176,0.501961,0.501961}
  \definecolor{LightCoral}{rgb}{0.941176,0.501961,0.501961}
  \definecolor{tomato}{rgb}{1.000000,0.388235,0.278431}
  \definecolor{orange red}{rgb}{1.000000,0.270588,0.000000}
  \definecolor{OrangeRed}{rgb}{1.000000,0.270588,0.000000}
  \definecolor{red}{rgb}{1.000000,0.000000,0.000000}
  \definecolor{hot pink}{rgb}{1.000000,0.411765,0.705882}
  \definecolor{HotPink}{rgb}{1.000000,0.411765,0.705882}
  \definecolor{deep pink}{rgb}{1.000000,0.078431,0.576471}
  \definecolor{DeepPink}{rgb}{1.000000,0.078431,0.576471}
  \definecolor{pink}{rgb}{1.000000,0.752941,0.796078}
  \definecolor{light pink}{rgb}{1.000000,0.713726,0.756863}
  \definecolor{LightPink}{rgb}{1.000000,0.713726,0.756863}
  \definecolor{pale violet red}{rgb}{0.858824,0.439216,0.576471}
  \definecolor{PaleVioletRed}{rgb}{0.858824,0.439216,0.576471}
  \definecolor{maroon}{rgb}{0.690196,0.188235,0.376471}
  \definecolor{medium violet red}{rgb}{0.780392,0.082353,0.521569}
  \definecolor{MediumVioletRed}{rgb}{0.780392,0.082353,0.521569}
  \definecolor{violet red}{rgb}{0.815686,0.125490,0.564706}
  \definecolor{VioletRed}{rgb}{0.815686,0.125490,0.564706}
  \definecolor{magenta}{rgb}{1.000000,0.000000,1.000000}
  \definecolor{violet}{rgb}{0.933333,0.509804,0.933333}
  \definecolor{plum}{rgb}{0.866667,0.627451,0.866667}
  \definecolor{orchid}{rgb}{0.854902,0.439216,0.839216}
  \definecolor{medium orchid}{rgb}{0.729412,0.333333,0.827451}
  \definecolor{MediumOrchid}{rgb}{0.729412,0.333333,0.827451}
  \definecolor{dark orchid}{rgb}{0.600000,0.196078,0.800000}
  \definecolor{DarkOrchid}{rgb}{0.600000,0.196078,0.800000}
  \definecolor{dark violet}{rgb}{0.580392,0.000000,0.827451}
  \definecolor{DarkViolet}{rgb}{0.580392,0.000000,0.827451}
  \definecolor{blue violet}{rgb}{0.541176,0.168627,0.886275}
  \definecolor{BlueViolet}{rgb}{0.541176,0.168627,0.886275}
  \definecolor{purple}{rgb}{0.627451,0.125490,0.941176}
  \definecolor{medium purple}{rgb}{0.576471,0.439216,0.858824}
  \definecolor{MediumPurple}{rgb}{0.576471,0.439216,0.858824}
  \definecolor{thistle}{rgb}{0.847059,0.749020,0.847059}
  \definecolor{snow1}{rgb}{1.000000,0.980392,0.980392}
  \definecolor{snow2}{rgb}{0.933333,0.913725,0.913725}
  \definecolor{snow3}{rgb}{0.803922,0.788235,0.788235}
  \definecolor{snow4}{rgb}{0.545098,0.537255,0.537255}
  \definecolor{seashell1}{rgb}{1.000000,0.960784,0.933333}
  \definecolor{seashell2}{rgb}{0.933333,0.898039,0.870588}
  \definecolor{seashell3}{rgb}{0.803922,0.772549,0.749020}
  \definecolor{seashell4}{rgb}{0.545098,0.525490,0.509804}
  \definecolor{AntiqueWhite1}{rgb}{1.000000,0.937255,0.858824}
  \definecolor{AntiqueWhite2}{rgb}{0.933333,0.874510,0.800000}
  \definecolor{AntiqueWhite3}{rgb}{0.803922,0.752941,0.690196}
  \definecolor{AntiqueWhite4}{rgb}{0.545098,0.513726,0.470588}
  \definecolor{bisque1}{rgb}{1.000000,0.894118,0.768627}
  \definecolor{bisque2}{rgb}{0.933333,0.835294,0.717647}
  \definecolor{bisque3}{rgb}{0.803922,0.717647,0.619608}
  \definecolor{bisque4}{rgb}{0.545098,0.490196,0.419608}
  \definecolor{PeachPuff1}{rgb}{1.000000,0.854902,0.725490}
  \definecolor{PeachPuff2}{rgb}{0.933333,0.796078,0.678431}
  \definecolor{PeachPuff3}{rgb}{0.803922,0.686275,0.584314}
  \definecolor{PeachPuff4}{rgb}{0.545098,0.466667,0.396078}
  \definecolor{NavajoWhite1}{rgb}{1.000000,0.870588,0.678431}
  \definecolor{NavajoWhite2}{rgb}{0.933333,0.811765,0.631373}
  \definecolor{NavajoWhite3}{rgb}{0.803922,0.701961,0.545098}
  \definecolor{NavajoWhite4}{rgb}{0.545098,0.474510,0.368627}
  \definecolor{LemonChiffon1}{rgb}{1.000000,0.980392,0.803922}
  \definecolor{LemonChiffon2}{rgb}{0.933333,0.913725,0.749020}
  \definecolor{LemonChiffon3}{rgb}{0.803922,0.788235,0.647059}
  \definecolor{LemonChiffon4}{rgb}{0.545098,0.537255,0.439216}
  \definecolor{cornsilk1}{rgb}{1.000000,0.972549,0.862745}
  \definecolor{cornsilk2}{rgb}{0.933333,0.909804,0.803922}
  \definecolor{cornsilk3}{rgb}{0.803922,0.784314,0.694118}
  \definecolor{cornsilk4}{rgb}{0.545098,0.533333,0.470588}
  \definecolor{ivory1}{rgb}{1.000000,1.000000,0.941176}
  \definecolor{ivory2}{rgb}{0.933333,0.933333,0.878431}
  \definecolor{ivory3}{rgb}{0.803922,0.803922,0.756863}
  \definecolor{ivory4}{rgb}{0.545098,0.545098,0.513726}
  \definecolor{honeydew1}{rgb}{0.941176,1.000000,0.941176}
  \definecolor{honeydew2}{rgb}{0.878431,0.933333,0.878431}
  \definecolor{honeydew3}{rgb}{0.756863,0.803922,0.756863}
  \definecolor{honeydew4}{rgb}{0.513726,0.545098,0.513726}
  \definecolor{LavenderBlush1}{rgb}{1.000000,0.941176,0.960784}
  \definecolor{LavenderBlush2}{rgb}{0.933333,0.878431,0.898039}
  \definecolor{LavenderBlush3}{rgb}{0.803922,0.756863,0.772549}
  \definecolor{LavenderBlush4}{rgb}{0.545098,0.513726,0.525490}
  \definecolor{MistyRose1}{rgb}{1.000000,0.894118,0.882353}
  \definecolor{MistyRose2}{rgb}{0.933333,0.835294,0.823529}
  \definecolor{MistyRose3}{rgb}{0.803922,0.717647,0.709804}
  \definecolor{MistyRose4}{rgb}{0.545098,0.490196,0.482353}
  \definecolor{azure1}{rgb}{0.941176,1.000000,1.000000}
  \definecolor{azure2}{rgb}{0.878431,0.933333,0.933333}
  \definecolor{azure3}{rgb}{0.756863,0.803922,0.803922}
  \definecolor{azure4}{rgb}{0.513726,0.545098,0.545098}
  \definecolor{SlateBlue1}{rgb}{0.513726,0.435294,1.000000}
  \definecolor{SlateBlue2}{rgb}{0.478431,0.403922,0.933333}
  \definecolor{SlateBlue3}{rgb}{0.411765,0.349020,0.803922}
  \definecolor{SlateBlue4}{rgb}{0.278431,0.235294,0.545098}
  \definecolor{RoyalBlue1}{rgb}{0.282353,0.462745,1.000000}
  \definecolor{RoyalBlue2}{rgb}{0.262745,0.431373,0.933333}
  \definecolor{RoyalBlue3}{rgb}{0.227451,0.372549,0.803922}
  \definecolor{RoyalBlue4}{rgb}{0.152941,0.250980,0.545098}
  \definecolor{blue1}{rgb}{0.000000,0.000000,1.000000}
  \definecolor{blue2}{rgb}{0.000000,0.000000,0.933333}
  \definecolor{blue3}{rgb}{0.000000,0.000000,0.803922}
  \definecolor{blue4}{rgb}{0.000000,0.000000,0.545098}
  \definecolor{DodgerBlue1}{rgb}{0.117647,0.564706,1.000000}
  \definecolor{DodgerBlue2}{rgb}{0.109804,0.525490,0.933333}
  \definecolor{DodgerBlue3}{rgb}{0.094118,0.454902,0.803922}
  \definecolor{DodgerBlue4}{rgb}{0.062745,0.305882,0.545098}
  \definecolor{SteelBlue1}{rgb}{0.388235,0.721569,1.000000}
  \definecolor{SteelBlue2}{rgb}{0.360784,0.674510,0.933333}
  \definecolor{SteelBlue3}{rgb}{0.309804,0.580392,0.803922}
  \definecolor{SteelBlue4}{rgb}{0.211765,0.392157,0.545098}
  \definecolor{DeepSkyBlue1}{rgb}{0.000000,0.749020,1.000000}
  \definecolor{DeepSkyBlue2}{rgb}{0.000000,0.698039,0.933333}
  \definecolor{DeepSkyBlue3}{rgb}{0.000000,0.603922,0.803922}
  \definecolor{DeepSkyBlue4}{rgb}{0.000000,0.407843,0.545098}
  \definecolor{SkyBlue1}{rgb}{0.529412,0.807843,1.000000}
  \definecolor{SkyBlue2}{rgb}{0.494118,0.752941,0.933333}
  \definecolor{SkyBlue3}{rgb}{0.423529,0.650980,0.803922}
  \definecolor{SkyBlue4}{rgb}{0.290196,0.439216,0.545098}
  \definecolor{LightSkyBlue1}{rgb}{0.690196,0.886275,1.000000}
  \definecolor{LightSkyBlue2}{rgb}{0.643137,0.827451,0.933333}
  \definecolor{LightSkyBlue3}{rgb}{0.552941,0.713726,0.803922}
  \definecolor{LightSkyBlue4}{rgb}{0.376471,0.482353,0.545098}
  \definecolor{SlateGray1}{rgb}{0.776471,0.886275,1.000000}
  \definecolor{SlateGray2}{rgb}{0.725490,0.827451,0.933333}
  \definecolor{SlateGray3}{rgb}{0.623529,0.713726,0.803922}
  \definecolor{SlateGray4}{rgb}{0.423529,0.482353,0.545098}
  \definecolor{LightSteelBlue1}{rgb}{0.792157,0.882353,1.000000}
  \definecolor{LightSteelBlue2}{rgb}{0.737255,0.823529,0.933333}
  \definecolor{LightSteelBlue3}{rgb}{0.635294,0.709804,0.803922}
  \definecolor{LightSteelBlue4}{rgb}{0.431373,0.482353,0.545098}
  \definecolor{LightBlue1}{rgb}{0.749020,0.937255,1.000000}
  \definecolor{LightBlue2}{rgb}{0.698039,0.874510,0.933333}
  \definecolor{LightBlue3}{rgb}{0.603922,0.752941,0.803922}
  \definecolor{LightBlue4}{rgb}{0.407843,0.513726,0.545098}
  \definecolor{LightCyan1}{rgb}{0.878431,1.000000,1.000000}
  \definecolor{LightCyan2}{rgb}{0.819608,0.933333,0.933333}
  \definecolor{LightCyan3}{rgb}{0.705882,0.803922,0.803922}
  \definecolor{LightCyan4}{rgb}{0.478431,0.545098,0.545098}
  \definecolor{PaleTurquoise1}{rgb}{0.733333,1.000000,1.000000}
  \definecolor{PaleTurquoise2}{rgb}{0.682353,0.933333,0.933333}
  \definecolor{PaleTurquoise3}{rgb}{0.588235,0.803922,0.803922}
  \definecolor{PaleTurquoise4}{rgb}{0.400000,0.545098,0.545098}
  \definecolor{CadetBlue1}{rgb}{0.596078,0.960784,1.000000}
  \definecolor{CadetBlue2}{rgb}{0.556863,0.898039,0.933333}
  \definecolor{CadetBlue3}{rgb}{0.478431,0.772549,0.803922}
  \definecolor{CadetBlue4}{rgb}{0.325490,0.525490,0.545098}
  \definecolor{turquoise1}{rgb}{0.000000,0.960784,1.000000}
  \definecolor{turquoise2}{rgb}{0.000000,0.898039,0.933333}
  \definecolor{turquoise3}{rgb}{0.000000,0.772549,0.803922}
  \definecolor{turquoise4}{rgb}{0.000000,0.525490,0.545098}
  \definecolor{cyan1}{rgb}{0.000000,1.000000,1.000000}
  \definecolor{cyan2}{rgb}{0.000000,0.933333,0.933333}
  \definecolor{cyan3}{rgb}{0.000000,0.803922,0.803922}
  \definecolor{cyan4}{rgb}{0.000000,0.545098,0.545098}
  \definecolor{DarkSlateGray1}{rgb}{0.592157,1.000000,1.000000}
  \definecolor{DarkSlateGray2}{rgb}{0.552941,0.933333,0.933333}
  \definecolor{DarkSlateGray3}{rgb}{0.474510,0.803922,0.803922}
  \definecolor{DarkSlateGray4}{rgb}{0.321569,0.545098,0.545098}
  \definecolor{aquamarine1}{rgb}{0.498039,1.000000,0.831373}
  \definecolor{aquamarine2}{rgb}{0.462745,0.933333,0.776471}
  \definecolor{aquamarine3}{rgb}{0.400000,0.803922,0.666667}
  \definecolor{aquamarine4}{rgb}{0.270588,0.545098,0.454902}
  \definecolor{DarkSeaGreen1}{rgb}{0.756863,1.000000,0.756863}
  \definecolor{DarkSeaGreen2}{rgb}{0.705882,0.933333,0.705882}
  \definecolor{DarkSeaGreen3}{rgb}{0.607843,0.803922,0.607843}
  \definecolor{DarkSeaGreen4}{rgb}{0.411765,0.545098,0.411765}
  \definecolor{SeaGreen1}{rgb}{0.329412,1.000000,0.623529}
  \definecolor{SeaGreen2}{rgb}{0.305882,0.933333,0.580392}
  \definecolor{SeaGreen3}{rgb}{0.262745,0.803922,0.501961}
  \definecolor{SeaGreen4}{rgb}{0.180392,0.545098,0.341176}
  \definecolor{PaleGreen1}{rgb}{0.603922,1.000000,0.603922}
  \definecolor{PaleGreen2}{rgb}{0.564706,0.933333,0.564706}
  \definecolor{PaleGreen3}{rgb}{0.486275,0.803922,0.486275}
  \definecolor{PaleGreen4}{rgb}{0.329412,0.545098,0.329412}
  \definecolor{SpringGreen1}{rgb}{0.000000,1.000000,0.498039}
  \definecolor{SpringGreen2}{rgb}{0.000000,0.933333,0.462745}
  \definecolor{SpringGreen3}{rgb}{0.000000,0.803922,0.400000}
  \definecolor{SpringGreen4}{rgb}{0.000000,0.545098,0.270588}
  \definecolor{green1}{rgb}{0.000000,1.000000,0.000000}
  \definecolor{green2}{rgb}{0.000000,0.933333,0.000000}
  \definecolor{green3}{rgb}{0.000000,0.803922,0.000000}
  \definecolor{green4}{rgb}{0.000000,0.545098,0.000000}
  \definecolor{chartreuse1}{rgb}{0.498039,1.000000,0.000000}
  \definecolor{chartreuse2}{rgb}{0.462745,0.933333,0.000000}
  \definecolor{chartreuse3}{rgb}{0.400000,0.803922,0.000000}
  \definecolor{chartreuse4}{rgb}{0.270588,0.545098,0.000000}
  \definecolor{OliveDrab1}{rgb}{0.752941,1.000000,0.243137}
  \definecolor{OliveDrab2}{rgb}{0.701961,0.933333,0.227451}
  \definecolor{OliveDrab3}{rgb}{0.603922,0.803922,0.196078}
  \definecolor{OliveDrab4}{rgb}{0.411765,0.545098,0.133333}
  \definecolor{DarkOliveGreen1}{rgb}{0.792157,1.000000,0.439216}
  \definecolor{DarkOliveGreen2}{rgb}{0.737255,0.933333,0.407843}
  \definecolor{DarkOliveGreen3}{rgb}{0.635294,0.803922,0.352941}
  \definecolor{DarkOliveGreen4}{rgb}{0.431373,0.545098,0.239216}
  \definecolor{khaki1}{rgb}{1.000000,0.964706,0.560784}
  \definecolor{khaki2}{rgb}{0.933333,0.901961,0.521569}
  \definecolor{khaki3}{rgb}{0.803922,0.776471,0.450980}
  \definecolor{khaki4}{rgb}{0.545098,0.525490,0.305882}
  \definecolor{LightGoldenrod1}{rgb}{1.000000,0.925490,0.545098}
  \definecolor{LightGoldenrod2}{rgb}{0.933333,0.862745,0.509804}
  \definecolor{LightGoldenrod3}{rgb}{0.803922,0.745098,0.439216}
  \definecolor{LightGoldenrod4}{rgb}{0.545098,0.505882,0.298039}
  \definecolor{LightYellow1}{rgb}{1.000000,1.000000,0.878431}
  \definecolor{LightYellow2}{rgb}{0.933333,0.933333,0.819608}
  \definecolor{LightYellow3}{rgb}{0.803922,0.803922,0.705882}
  \definecolor{LightYellow4}{rgb}{0.545098,0.545098,0.478431}
  \definecolor{yellow1}{rgb}{1.000000,1.000000,0.000000}
  \definecolor{yellow2}{rgb}{0.933333,0.933333,0.000000}
  \definecolor{yellow3}{rgb}{0.803922,0.803922,0.000000}
  \definecolor{yellow4}{rgb}{0.545098,0.545098,0.000000}
  \definecolor{gold1}{rgb}{1.000000,0.843137,0.000000}
  \definecolor{gold2}{rgb}{0.933333,0.788235,0.000000}
  \definecolor{gold3}{rgb}{0.803922,0.678431,0.000000}
  \definecolor{gold4}{rgb}{0.545098,0.458824,0.000000}
  \definecolor{goldenrod1}{rgb}{1.000000,0.756863,0.145098}
  \definecolor{goldenrod2}{rgb}{0.933333,0.705882,0.133333}
  \definecolor{goldenrod3}{rgb}{0.803922,0.607843,0.113725}
  \definecolor{goldenrod4}{rgb}{0.545098,0.411765,0.078431}
  \definecolor{DarkGoldenrod1}{rgb}{1.000000,0.725490,0.058824}
  \definecolor{DarkGoldenrod2}{rgb}{0.933333,0.678431,0.054902}
  \definecolor{DarkGoldenrod3}{rgb}{0.803922,0.584314,0.047059}
  \definecolor{DarkGoldenrod4}{rgb}{0.545098,0.396078,0.031373}
  \definecolor{RosyBrown1}{rgb}{1.000000,0.756863,0.756863}
  \definecolor{RosyBrown2}{rgb}{0.933333,0.705882,0.705882}
  \definecolor{RosyBrown3}{rgb}{0.803922,0.607843,0.607843}
  \definecolor{RosyBrown4}{rgb}{0.545098,0.411765,0.411765}
  \definecolor{IndianRed1}{rgb}{1.000000,0.415686,0.415686}
  \definecolor{IndianRed2}{rgb}{0.933333,0.388235,0.388235}
  \definecolor{IndianRed3}{rgb}{0.803922,0.333333,0.333333}
  \definecolor{IndianRed4}{rgb}{0.545098,0.227451,0.227451}
  \definecolor{sienna1}{rgb}{1.000000,0.509804,0.278431}
  \definecolor{sienna2}{rgb}{0.933333,0.474510,0.258824}
  \definecolor{sienna3}{rgb}{0.803922,0.407843,0.223529}
  \definecolor{sienna4}{rgb}{0.545098,0.278431,0.149020}
  \definecolor{burlywood1}{rgb}{1.000000,0.827451,0.607843}
  \definecolor{burlywood2}{rgb}{0.933333,0.772549,0.568627}
  \definecolor{burlywood3}{rgb}{0.803922,0.666667,0.490196}
  \definecolor{burlywood4}{rgb}{0.545098,0.450980,0.333333}
  \definecolor{wheat1}{rgb}{1.000000,0.905882,0.729412}
  \definecolor{wheat2}{rgb}{0.933333,0.847059,0.682353}
  \definecolor{wheat3}{rgb}{0.803922,0.729412,0.588235}
  \definecolor{wheat4}{rgb}{0.545098,0.494118,0.400000}
  \definecolor{tan1}{rgb}{1.000000,0.647059,0.309804}
  \definecolor{tan2}{rgb}{0.933333,0.603922,0.286275}
  \definecolor{tan3}{rgb}{0.803922,0.521569,0.247059}
  \definecolor{tan4}{rgb}{0.545098,0.352941,0.168627}
  \definecolor{chocolate1}{rgb}{1.000000,0.498039,0.141176}
  \definecolor{chocolate2}{rgb}{0.933333,0.462745,0.129412}
  \definecolor{chocolate3}{rgb}{0.803922,0.400000,0.113725}
  \definecolor{chocolate4}{rgb}{0.545098,0.270588,0.074510}
  \definecolor{firebrick1}{rgb}{1.000000,0.188235,0.188235}
  \definecolor{firebrick2}{rgb}{0.933333,0.172549,0.172549}
  \definecolor{firebrick3}{rgb}{0.803922,0.149020,0.149020}
  \definecolor{firebrick4}{rgb}{0.545098,0.101961,0.101961}
  \definecolor{brown1}{rgb}{1.000000,0.250980,0.250980}
  \definecolor{brown2}{rgb}{0.933333,0.231373,0.231373}
  \definecolor{brown3}{rgb}{0.803922,0.200000,0.200000}
  \definecolor{brown4}{rgb}{0.545098,0.137255,0.137255}
  \definecolor{salmon1}{rgb}{1.000000,0.549020,0.411765}
  \definecolor{salmon2}{rgb}{0.933333,0.509804,0.384314}
  \definecolor{salmon3}{rgb}{0.803922,0.439216,0.329412}
  \definecolor{salmon4}{rgb}{0.545098,0.298039,0.223529}
  \definecolor{LightSalmon1}{rgb}{1.000000,0.627451,0.478431}
  \definecolor{LightSalmon2}{rgb}{0.933333,0.584314,0.447059}
  \definecolor{LightSalmon3}{rgb}{0.803922,0.505882,0.384314}
  \definecolor{LightSalmon4}{rgb}{0.545098,0.341176,0.258824}
  \definecolor{orange1}{rgb}{1.000000,0.647059,0.000000}
  \definecolor{orange2}{rgb}{0.933333,0.603922,0.000000}
  \definecolor{orange3}{rgb}{0.803922,0.521569,0.000000}
  \definecolor{orange4}{rgb}{0.545098,0.352941,0.000000}
  \definecolor{DarkOrange1}{rgb}{1.000000,0.498039,0.000000}
  \definecolor{DarkOrange2}{rgb}{0.933333,0.462745,0.000000}
  \definecolor{DarkOrange3}{rgb}{0.803922,0.400000,0.000000}
  \definecolor{DarkOrange4}{rgb}{0.545098,0.270588,0.000000}
  \definecolor{coral1}{rgb}{1.000000,0.447059,0.337255}
  \definecolor{coral2}{rgb}{0.933333,0.415686,0.313726}
  \definecolor{coral3}{rgb}{0.803922,0.356863,0.270588}
  \definecolor{coral4}{rgb}{0.545098,0.243137,0.184314}
  \definecolor{tomato1}{rgb}{1.000000,0.388235,0.278431}
  \definecolor{tomato2}{rgb}{0.933333,0.360784,0.258824}
  \definecolor{tomato3}{rgb}{0.803922,0.309804,0.223529}
  \definecolor{tomato4}{rgb}{0.545098,0.211765,0.149020}
  \definecolor{OrangeRed1}{rgb}{1.000000,0.270588,0.000000}
  \definecolor{OrangeRed2}{rgb}{0.933333,0.250980,0.000000}
  \definecolor{OrangeRed3}{rgb}{0.803922,0.215686,0.000000}
  \definecolor{OrangeRed4}{rgb}{0.545098,0.145098,0.000000}
  \definecolor{red1}{rgb}{1.000000,0.000000,0.000000}
  \definecolor{red2}{rgb}{0.933333,0.000000,0.000000}
  \definecolor{red3}{rgb}{0.803922,0.000000,0.000000}
  \definecolor{red4}{rgb}{0.545098,0.000000,0.000000}
  \definecolor{DeepPink1}{rgb}{1.000000,0.078431,0.576471}
  \definecolor{DeepPink2}{rgb}{0.933333,0.070588,0.537255}
  \definecolor{DeepPink3}{rgb}{0.803922,0.062745,0.462745}
  \definecolor{DeepPink4}{rgb}{0.545098,0.039216,0.313726}
  \definecolor{HotPink1}{rgb}{1.000000,0.431373,0.705882}
  \definecolor{HotPink2}{rgb}{0.933333,0.415686,0.654902}
  \definecolor{HotPink3}{rgb}{0.803922,0.376471,0.564706}
  \definecolor{HotPink4}{rgb}{0.545098,0.227451,0.384314}
  \definecolor{pink1}{rgb}{1.000000,0.709804,0.772549}
  \definecolor{pink2}{rgb}{0.933333,0.662745,0.721569}
  \definecolor{pink3}{rgb}{0.803922,0.568627,0.619608}
  \definecolor{pink4}{rgb}{0.545098,0.388235,0.423529}
  \definecolor{LightPink1}{rgb}{1.000000,0.682353,0.725490}
  \definecolor{LightPink2}{rgb}{0.933333,0.635294,0.678431}
  \definecolor{LightPink3}{rgb}{0.803922,0.549020,0.584314}
  \definecolor{LightPink4}{rgb}{0.545098,0.372549,0.396078}
  \definecolor{PaleVioletRed1}{rgb}{1.000000,0.509804,0.670588}
  \definecolor{PaleVioletRed2}{rgb}{0.933333,0.474510,0.623529}
  \definecolor{PaleVioletRed3}{rgb}{0.803922,0.407843,0.537255}
  \definecolor{PaleVioletRed4}{rgb}{0.545098,0.278431,0.364706}
  \definecolor{maroon1}{rgb}{1.000000,0.203922,0.701961}
  \definecolor{maroon2}{rgb}{0.933333,0.188235,0.654902}
  \definecolor{maroon3}{rgb}{0.803922,0.160784,0.564706}
  \definecolor{maroon4}{rgb}{0.545098,0.109804,0.384314}
  \definecolor{VioletRed1}{rgb}{1.000000,0.243137,0.588235}
  \definecolor{VioletRed2}{rgb}{0.933333,0.227451,0.549020}
  \definecolor{VioletRed3}{rgb}{0.803922,0.196078,0.470588}
  \definecolor{VioletRed4}{rgb}{0.545098,0.133333,0.321569}
  \definecolor{magenta1}{rgb}{1.000000,0.000000,1.000000}
  \definecolor{magenta2}{rgb}{0.933333,0.000000,0.933333}
  \definecolor{magenta3}{rgb}{0.803922,0.000000,0.803922}
  \definecolor{magenta4}{rgb}{0.545098,0.000000,0.545098}
  \definecolor{orchid1}{rgb}{1.000000,0.513726,0.980392}
  \definecolor{orchid2}{rgb}{0.933333,0.478431,0.913725}
  \definecolor{orchid3}{rgb}{0.803922,0.411765,0.788235}
  \definecolor{orchid4}{rgb}{0.545098,0.278431,0.537255}
  \definecolor{plum1}{rgb}{1.000000,0.733333,1.000000}
  \definecolor{plum2}{rgb}{0.933333,0.682353,0.933333}
  \definecolor{plum3}{rgb}{0.803922,0.588235,0.803922}
  \definecolor{plum4}{rgb}{0.545098,0.400000,0.545098}
  \definecolor{MediumOrchid1}{rgb}{0.878431,0.400000,1.000000}
  \definecolor{MediumOrchid2}{rgb}{0.819608,0.372549,0.933333}
  \definecolor{MediumOrchid3}{rgb}{0.705882,0.321569,0.803922}
  \definecolor{MediumOrchid4}{rgb}{0.478431,0.215686,0.545098}
  \definecolor{DarkOrchid1}{rgb}{0.749020,0.243137,1.000000}
  \definecolor{DarkOrchid2}{rgb}{0.698039,0.227451,0.933333}
  \definecolor{DarkOrchid3}{rgb}{0.603922,0.196078,0.803922}
  \definecolor{DarkOrchid4}{rgb}{0.407843,0.133333,0.545098}
  \definecolor{purple1}{rgb}{0.607843,0.188235,1.000000}
  \definecolor{purple2}{rgb}{0.568627,0.172549,0.933333}
  \definecolor{purple3}{rgb}{0.490196,0.149020,0.803922}
  \definecolor{purple4}{rgb}{0.333333,0.101961,0.545098}
  \definecolor{MediumPurple1}{rgb}{0.670588,0.509804,1.000000}
  \definecolor{MediumPurple2}{rgb}{0.623529,0.474510,0.933333}
  \definecolor{MediumPurple3}{rgb}{0.537255,0.407843,0.803922}
  \definecolor{MediumPurple4}{rgb}{0.364706,0.278431,0.545098}
  \definecolor{thistle1}{rgb}{1.000000,0.882353,1.000000}
  \definecolor{thistle2}{rgb}{0.933333,0.823529,0.933333}
  \definecolor{thistle3}{rgb}{0.803922,0.709804,0.803922}
  \definecolor{thistle4}{rgb}{0.545098,0.482353,0.545098}
  \definecolor{gray0}{rgb}{0.000000,0.000000,0.000000}
  \definecolor{grey0}{rgb}{0.000000,0.000000,0.000000}
  \definecolor{gray1}{rgb}{0.011765,0.011765,0.011765}
  \definecolor{grey1}{rgb}{0.011765,0.011765,0.011765}
  \definecolor{gray2}{rgb}{0.019608,0.019608,0.019608}
  \definecolor{grey2}{rgb}{0.019608,0.019608,0.019608}
  \definecolor{gray3}{rgb}{0.031373,0.031373,0.031373}
  \definecolor{grey3}{rgb}{0.031373,0.031373,0.031373}
  \definecolor{gray4}{rgb}{0.039216,0.039216,0.039216}
  \definecolor{grey4}{rgb}{0.039216,0.039216,0.039216}
  \definecolor{gray5}{rgb}{0.050980,0.050980,0.050980}
  \definecolor{grey5}{rgb}{0.050980,0.050980,0.050980}
  \definecolor{gray6}{rgb}{0.058824,0.058824,0.058824}
  \definecolor{grey6}{rgb}{0.058824,0.058824,0.058824}
  \definecolor{gray7}{rgb}{0.070588,0.070588,0.070588}
  \definecolor{grey7}{rgb}{0.070588,0.070588,0.070588}
  \definecolor{gray8}{rgb}{0.078431,0.078431,0.078431}
  \definecolor{grey8}{rgb}{0.078431,0.078431,0.078431}
  \definecolor{gray9}{rgb}{0.090196,0.090196,0.090196}
  \definecolor{grey9}{rgb}{0.090196,0.090196,0.090196}
  \definecolor{gray10}{rgb}{0.101961,0.101961,0.101961}
  \definecolor{grey10}{rgb}{0.101961,0.101961,0.101961}
  \definecolor{gray11}{rgb}{0.109804,0.109804,0.109804}
  \definecolor{grey11}{rgb}{0.109804,0.109804,0.109804}
  \definecolor{gray12}{rgb}{0.121569,0.121569,0.121569}
  \definecolor{grey12}{rgb}{0.121569,0.121569,0.121569}
  \definecolor{gray13}{rgb}{0.129412,0.129412,0.129412}
  \definecolor{grey13}{rgb}{0.129412,0.129412,0.129412}
  \definecolor{gray14}{rgb}{0.141176,0.141176,0.141176}
  \definecolor{grey14}{rgb}{0.141176,0.141176,0.141176}
  \definecolor{gray15}{rgb}{0.149020,0.149020,0.149020}
  \definecolor{grey15}{rgb}{0.149020,0.149020,0.149020}
  \definecolor{gray16}{rgb}{0.160784,0.160784,0.160784}
  \definecolor{grey16}{rgb}{0.160784,0.160784,0.160784}
  \definecolor{gray17}{rgb}{0.168627,0.168627,0.168627}
  \definecolor{grey17}{rgb}{0.168627,0.168627,0.168627}
  \definecolor{gray18}{rgb}{0.180392,0.180392,0.180392}
  \definecolor{grey18}{rgb}{0.180392,0.180392,0.180392}
  \definecolor{gray19}{rgb}{0.188235,0.188235,0.188235}
  \definecolor{grey19}{rgb}{0.188235,0.188235,0.188235}
  \definecolor{gray20}{rgb}{0.200000,0.200000,0.200000}
  \definecolor{grey20}{rgb}{0.200000,0.200000,0.200000}
  \definecolor{gray21}{rgb}{0.211765,0.211765,0.211765}
  \definecolor{grey21}{rgb}{0.211765,0.211765,0.211765}
  \definecolor{gray22}{rgb}{0.219608,0.219608,0.219608}
  \definecolor{grey22}{rgb}{0.219608,0.219608,0.219608}
  \definecolor{gray23}{rgb}{0.231373,0.231373,0.231373}
  \definecolor{grey23}{rgb}{0.231373,0.231373,0.231373}
  \definecolor{gray24}{rgb}{0.239216,0.239216,0.239216}
  \definecolor{grey24}{rgb}{0.239216,0.239216,0.239216}
  \definecolor{gray25}{rgb}{0.250980,0.250980,0.250980}
  \definecolor{grey25}{rgb}{0.250980,0.250980,0.250980}
  \definecolor{gray26}{rgb}{0.258824,0.258824,0.258824}
  \definecolor{grey26}{rgb}{0.258824,0.258824,0.258824}
  \definecolor{gray27}{rgb}{0.270588,0.270588,0.270588}
  \definecolor{grey27}{rgb}{0.270588,0.270588,0.270588}
  \definecolor{gray28}{rgb}{0.278431,0.278431,0.278431}
  \definecolor{grey28}{rgb}{0.278431,0.278431,0.278431}
  \definecolor{gray29}{rgb}{0.290196,0.290196,0.290196}
  \definecolor{grey29}{rgb}{0.290196,0.290196,0.290196}
  \definecolor{gray30}{rgb}{0.301961,0.301961,0.301961}
  \definecolor{grey30}{rgb}{0.301961,0.301961,0.301961}
  \definecolor{gray31}{rgb}{0.309804,0.309804,0.309804}
  \definecolor{grey31}{rgb}{0.309804,0.309804,0.309804}
  \definecolor{gray32}{rgb}{0.321569,0.321569,0.321569}
  \definecolor{grey32}{rgb}{0.321569,0.321569,0.321569}
  \definecolor{gray33}{rgb}{0.329412,0.329412,0.329412}
  \definecolor{grey33}{rgb}{0.329412,0.329412,0.329412}
  \definecolor{gray34}{rgb}{0.341176,0.341176,0.341176}
  \definecolor{grey34}{rgb}{0.341176,0.341176,0.341176}
  \definecolor{gray35}{rgb}{0.349020,0.349020,0.349020}
  \definecolor{grey35}{rgb}{0.349020,0.349020,0.349020}
  \definecolor{gray36}{rgb}{0.360784,0.360784,0.360784}
  \definecolor{grey36}{rgb}{0.360784,0.360784,0.360784}
  \definecolor{gray37}{rgb}{0.368627,0.368627,0.368627}
  \definecolor{grey37}{rgb}{0.368627,0.368627,0.368627}
  \definecolor{gray38}{rgb}{0.380392,0.380392,0.380392}
  \definecolor{grey38}{rgb}{0.380392,0.380392,0.380392}
  \definecolor{gray39}{rgb}{0.388235,0.388235,0.388235}
  \definecolor{grey39}{rgb}{0.388235,0.388235,0.388235}
  \definecolor{gray40}{rgb}{0.400000,0.400000,0.400000}
  \definecolor{grey40}{rgb}{0.400000,0.400000,0.400000}
  \definecolor{gray41}{rgb}{0.411765,0.411765,0.411765}
  \definecolor{grey41}{rgb}{0.411765,0.411765,0.411765}
  \definecolor{gray42}{rgb}{0.419608,0.419608,0.419608}
  \definecolor{grey42}{rgb}{0.419608,0.419608,0.419608}
  \definecolor{gray43}{rgb}{0.431373,0.431373,0.431373}
  \definecolor{grey43}{rgb}{0.431373,0.431373,0.431373}
  \definecolor{gray44}{rgb}{0.439216,0.439216,0.439216}
  \definecolor{grey44}{rgb}{0.439216,0.439216,0.439216}
  \definecolor{gray45}{rgb}{0.450980,0.450980,0.450980}
  \definecolor{grey45}{rgb}{0.450980,0.450980,0.450980}
  \definecolor{gray46}{rgb}{0.458824,0.458824,0.458824}
  \definecolor{grey46}{rgb}{0.458824,0.458824,0.458824}
  \definecolor{gray47}{rgb}{0.470588,0.470588,0.470588}
  \definecolor{grey47}{rgb}{0.470588,0.470588,0.470588}
  \definecolor{gray48}{rgb}{0.478431,0.478431,0.478431}
  \definecolor{grey48}{rgb}{0.478431,0.478431,0.478431}
  \definecolor{gray49}{rgb}{0.490196,0.490196,0.490196}
  \definecolor{grey49}{rgb}{0.490196,0.490196,0.490196}
  \definecolor{gray50}{rgb}{0.498039,0.498039,0.498039}
  \definecolor{grey50}{rgb}{0.498039,0.498039,0.498039}
  \definecolor{gray51}{rgb}{0.509804,0.509804,0.509804}
  \definecolor{grey51}{rgb}{0.509804,0.509804,0.509804}
  \definecolor{gray52}{rgb}{0.521569,0.521569,0.521569}
  \definecolor{grey52}{rgb}{0.521569,0.521569,0.521569}
  \definecolor{gray53}{rgb}{0.529412,0.529412,0.529412}
  \definecolor{grey53}{rgb}{0.529412,0.529412,0.529412}
  \definecolor{gray54}{rgb}{0.541176,0.541176,0.541176}
  \definecolor{grey54}{rgb}{0.541176,0.541176,0.541176}
  \definecolor{gray55}{rgb}{0.549020,0.549020,0.549020}
  \definecolor{grey55}{rgb}{0.549020,0.549020,0.549020}
  \definecolor{gray56}{rgb}{0.560784,0.560784,0.560784}
  \definecolor{grey56}{rgb}{0.560784,0.560784,0.560784}
  \definecolor{gray57}{rgb}{0.568627,0.568627,0.568627}
  \definecolor{grey57}{rgb}{0.568627,0.568627,0.568627}
  \definecolor{gray58}{rgb}{0.580392,0.580392,0.580392}
  \definecolor{grey58}{rgb}{0.580392,0.580392,0.580392}
  \definecolor{gray59}{rgb}{0.588235,0.588235,0.588235}
  \definecolor{grey59}{rgb}{0.588235,0.588235,0.588235}
  \definecolor{gray60}{rgb}{0.600000,0.600000,0.600000}
  \definecolor{grey60}{rgb}{0.600000,0.600000,0.600000}
  \definecolor{gray61}{rgb}{0.611765,0.611765,0.611765}
  \definecolor{grey61}{rgb}{0.611765,0.611765,0.611765}
  \definecolor{gray62}{rgb}{0.619608,0.619608,0.619608}
  \definecolor{grey62}{rgb}{0.619608,0.619608,0.619608}
  \definecolor{gray63}{rgb}{0.631373,0.631373,0.631373}
  \definecolor{grey63}{rgb}{0.631373,0.631373,0.631373}
  \definecolor{gray64}{rgb}{0.639216,0.639216,0.639216}
  \definecolor{grey64}{rgb}{0.639216,0.639216,0.639216}
  \definecolor{gray65}{rgb}{0.650980,0.650980,0.650980}
  \definecolor{grey65}{rgb}{0.650980,0.650980,0.650980}
  \definecolor{gray66}{rgb}{0.658824,0.658824,0.658824}
  \definecolor{grey66}{rgb}{0.658824,0.658824,0.658824}
  \definecolor{gray67}{rgb}{0.670588,0.670588,0.670588}
  \definecolor{grey67}{rgb}{0.670588,0.670588,0.670588}
  \definecolor{gray68}{rgb}{0.678431,0.678431,0.678431}
  \definecolor{grey68}{rgb}{0.678431,0.678431,0.678431}
  \definecolor{gray69}{rgb}{0.690196,0.690196,0.690196}
  \definecolor{grey69}{rgb}{0.690196,0.690196,0.690196}
  \definecolor{gray70}{rgb}{0.701961,0.701961,0.701961}
  \definecolor{grey70}{rgb}{0.701961,0.701961,0.701961}
  \definecolor{gray71}{rgb}{0.709804,0.709804,0.709804}
  \definecolor{grey71}{rgb}{0.709804,0.709804,0.709804}
  \definecolor{gray72}{rgb}{0.721569,0.721569,0.721569}
  \definecolor{grey72}{rgb}{0.721569,0.721569,0.721569}
  \definecolor{gray73}{rgb}{0.729412,0.729412,0.729412}
  \definecolor{grey73}{rgb}{0.729412,0.729412,0.729412}
  \definecolor{gray74}{rgb}{0.741176,0.741176,0.741176}
  \definecolor{grey74}{rgb}{0.741176,0.741176,0.741176}
  \definecolor{gray75}{rgb}{0.749020,0.749020,0.749020}
  \definecolor{grey75}{rgb}{0.749020,0.749020,0.749020}
  \definecolor{gray76}{rgb}{0.760784,0.760784,0.760784}
  \definecolor{grey76}{rgb}{0.760784,0.760784,0.760784}
  \definecolor{gray77}{rgb}{0.768627,0.768627,0.768627}
  \definecolor{grey77}{rgb}{0.768627,0.768627,0.768627}
  \definecolor{gray78}{rgb}{0.780392,0.780392,0.780392}
  \definecolor{grey78}{rgb}{0.780392,0.780392,0.780392}
  \definecolor{gray79}{rgb}{0.788235,0.788235,0.788235}
  \definecolor{grey79}{rgb}{0.788235,0.788235,0.788235}
  \definecolor{gray80}{rgb}{0.800000,0.800000,0.800000}
  \definecolor{grey80}{rgb}{0.800000,0.800000,0.800000}
  \definecolor{gray81}{rgb}{0.811765,0.811765,0.811765}
  \definecolor{grey81}{rgb}{0.811765,0.811765,0.811765}
  \definecolor{gray82}{rgb}{0.819608,0.819608,0.819608}
  \definecolor{grey82}{rgb}{0.819608,0.819608,0.819608}
  \definecolor{gray83}{rgb}{0.831373,0.831373,0.831373}
  \definecolor{grey83}{rgb}{0.831373,0.831373,0.831373}
  \definecolor{gray84}{rgb}{0.839216,0.839216,0.839216}
  \definecolor{grey84}{rgb}{0.839216,0.839216,0.839216}
  \definecolor{gray85}{rgb}{0.850980,0.850980,0.850980}
  \definecolor{grey85}{rgb}{0.850980,0.850980,0.850980}
  \definecolor{gray86}{rgb}{0.858824,0.858824,0.858824}
  \definecolor{grey86}{rgb}{0.858824,0.858824,0.858824}
  \definecolor{gray87}{rgb}{0.870588,0.870588,0.870588}
  \definecolor{grey87}{rgb}{0.870588,0.870588,0.870588}
  \definecolor{gray88}{rgb}{0.878431,0.878431,0.878431}
  \definecolor{grey88}{rgb}{0.878431,0.878431,0.878431}
  \definecolor{gray89}{rgb}{0.890196,0.890196,0.890196}
  \definecolor{grey89}{rgb}{0.890196,0.890196,0.890196}
  \definecolor{gray90}{rgb}{0.898039,0.898039,0.898039}
  \definecolor{grey90}{rgb}{0.898039,0.898039,0.898039}
  \definecolor{gray91}{rgb}{0.909804,0.909804,0.909804}
  \definecolor{grey91}{rgb}{0.909804,0.909804,0.909804}
  \definecolor{gray92}{rgb}{0.921569,0.921569,0.921569}
  \definecolor{grey92}{rgb}{0.921569,0.921569,0.921569}
  \definecolor{gray93}{rgb}{0.929412,0.929412,0.929412}
  \definecolor{grey93}{rgb}{0.929412,0.929412,0.929412}
  \definecolor{gray94}{rgb}{0.941176,0.941176,0.941176}
  \definecolor{grey94}{rgb}{0.941176,0.941176,0.941176}
  \definecolor{gray95}{rgb}{0.949020,0.949020,0.949020}
  \definecolor{grey95}{rgb}{0.949020,0.949020,0.949020}
  \definecolor{gray96}{rgb}{0.960784,0.960784,0.960784}
  \definecolor{grey96}{rgb}{0.960784,0.960784,0.960784}
  \definecolor{gray97}{rgb}{0.968627,0.968627,0.968627}
  \definecolor{grey97}{rgb}{0.968627,0.968627,0.968627}
  \definecolor{gray98}{rgb}{0.980392,0.980392,0.980392}
  \definecolor{grey98}{rgb}{0.980392,0.980392,0.980392}
  \definecolor{gray99}{rgb}{0.988235,0.988235,0.988235}
  \definecolor{grey99}{rgb}{0.988235,0.988235,0.988235}
  \definecolor{gray100}{rgb}{1.000000,1.000000,1.000000}
  \definecolor{grey100}{rgb}{1.000000,1.000000,1.000000}
  \definecolor{dark grey}{rgb}{0.662745,0.662745,0.662745}
  \definecolor{DarkGrey}{rgb}{0.662745,0.662745,0.662745}
  \definecolor{dark gray}{rgb}{0.662745,0.662745,0.662745}
  \definecolor{DarkGray}{rgb}{0.662745,0.662745,0.662745}
  \definecolor{dark blue}{rgb}{0.000000,0.000000,0.545098}
  \definecolor{DarkBlue}{rgb}{0.000000,0.000000,0.545098}
  \definecolor{dark cyan}{rgb}{0.000000,0.545098,0.545098}
  \definecolor{DarkCyan}{rgb}{0.000000,0.545098,0.545098}
  \definecolor{dark magenta}{rgb}{0.545098,0.000000,0.545098}
  \definecolor{DarkMagenta}{rgb}{0.545098,0.000000,0.545098}
  \definecolor{dark red}{rgb}{0.545098,0.000000,0.000000}
  \definecolor{DarkRed}{rgb}{0.545098,0.000000,0.000000}
  \definecolor{light green}{rgb}{0.564706,0.933333,0.564706}
  \definecolor{LightGreen}{rgb}{0.564706,0.933333,0.564706}

\usepackage{etex}
\reserveinserts{100}
\usepackage{morefloats}
\usepackage{paralist}
\usepackage{xspace}
\usepackage{tabularx}
\usepackage{booktabs}
\usepackage{wrapfig}

\usepackage{mathdots}
\usepackage{lscape}

\usepackage{listings}

\usepackage{algorithm}
\usepackage{algorithmicx}
\usepackage{algpseudocode}
\algnotext{EndFor}
\algnotext{EndIf}
\algnotext{EndProcedure}

\newlength{\arrayrulewidthOriginal}
\newcommand{\Cline}[2]{
  \noalign{\global\setlength{\arrayrulewidthOriginal}{\arrayrulewidth}}
  \noalign{\global\setlength{\arrayrulewidth}{#1}}\cline{#2}
  \noalign{\global\setlength{\arrayrulewidth}{\arrayrulewidthOriginal}}}

\newcommand{\algrule}[1][.2pt]{\par\vskip.5\baselineskip\hrule height #1\par\vskip.5\baselineskip} 

\providecommand{\gG}{\ensuremath{G}}
\providecommand{\gH}{\ensuremath{H}}

\providecommand{\fX}{\ensuremath{\mX}}

\providecommand{\ops}{\ensuremath{\Phi}}
\providecommand{\primops}{\ensuremath{P}}

\providecommand{\fset}{\ensuremath{F}}

\providecommand{\simfun}{\ensuremath{\mathsf{S}}}

\renewcommand{\E}{\ensuremath{{E}}}
\newcommand{\X}{\ensuremath{\mX}}

\renewcommand{\H}{\ensuremath{\mH}}

\newcommand{\N}{\ensuremath{\mathcal{N}}}

\newcommand{\simmat}{\ensuremath{\mathbf{S}}}
\renewcommand{\sim}{\ensuremath{\mathsf{S}}}

\newcommand{\maxiter}{\ensuremath{\mathsf{maxiter}}}

\definecolor{plotblue}{RGB}	{30,144,255}
\definecolor{plotgreen}{RGB}	{50,205,50}
\definecolor{plotred}{RGB}	{220,20,60}

\definecolor{myyellow}{RGB}{255,255,204}
\definecolor{myred}{RGB}{255,204,204}
\definecolor{myblue}{RGB}{0,200,255}
\definecolor{mygreen}{RGB}{80,220,80}

\newcommand*\hrulefillvar[1][0.4pt]{\leavevmode\leaders\hrule height#1\hfill\kern0pt}
\newcommand{\etal}{\emph{et al.}\xspace}

\newcommand\T{\rule{0pt}{3.2ex}}
\newcommand\B{\rule[-1.4ex]{0pt}{0pt}}

\newcommand\TT{\rule{0pt}{2.3ex}}
\newcommand\BB{\rule[-1.0ex]{0pt}{0pt}}

\newcommand\TTZ{\rule{0pt}{2.6ex}}
\newcommand\BBZ{\rule[-1.4ex]{0pt}{0pt}}

\definecolor{gray}{RGB}{20,20,20}
\definecolor{greencm}{RGB}{0,153,0}

\usepackage{amsthm}

\newtheoremstyle{probstyle}               
  {1.4mm}
  {1.4mm}
  {\itshape}
  {}
  {\bfseries}
  {:}
  { }
  {}
\theoremstyle{probstyle}

\newtheoremstyle{mystyle}
  {1.5mm}
  {1.5mm}
  {\itshape}
  {}
  {\bfseries}
  {:}
  { }
  {}
\theoremstyle{mystyle}

\newtheorem{Definition}{Definition}[section]

\newtheoremstyle{newstyle}
  {1.0mm}
  {1.0mm}
  {\itshape}
  {}
  {\sf \it \bfseries}
  {:}
  { }
  {}
\theoremstyle{newstyle}

\newtheoremstyle{nonumberplain}
  {-0.2}
  {}
  {\normalfont\upshape}
  {1mm}
  {\normalfont\small\sffamily\sc}
  {:}
  { }
  {}
\theoremstyle{nonumberplain}

\newcolumntype{H}{>{\setbox0=\hbox\bgroup}c<{\egroup}@{}}

\definecolor{gray}{RGB}{150,150,150}
\definecolor{theblue}{RGB}{0,0,180}
\usepackage{hyperref}
\hypersetup{colorlinks=true,urlcolor=theblue,citecolor=theblue,linkcolor=theblue}

\hypersetup{
pdftitle={Role Discovery in Networks},
pdfauthor={Ryan A. Rossi and Nesreen K. Ahmed},
}   

\ifCLASSOPTIONcompsoc
\else
\fi

\newcommand\ZZ{\rule{0pt}{2.6ex}}
\newcommand\XX{\rule[-1.2ex]{0pt}{0pt}}

\usepackage{flushend}
\usepackage{url}
\usepackage{amssymb}
\usepackage{graphicx}

\begin{document}
\title{Role Discovery in Networks}

\author{Ryan A. Rossi and Nesreen K. Ahmed
\IEEEcompsocitemizethanks{\IEEEcompsocthanksitem R. A. Rossi and N. K. Ahmed are with the Department
of Computer Science, Purdue University, West Lafayette, IN, 47907.\protect\\
E-mail: rrossi@purdue.edu, nkahmed@purdue.edu
}
\thanks{}}

\markboth{
}
{Rossi \MakeLowercase{\textit{et al.}}: Role Discovery in Networks}

\IEEEcompsoctitleabstractindextext{
\begin{abstract}
Roles represent node-level connectivity patterns such as star-center, star-edge nodes, near-cliques or nodes that act as bridges to different regions of the graph.
Intuitively, two nodes belong to the same role if they are structurally similar.
Roles have been mainly of interest to sociologists, but more recently, roles have become increasingly useful in other domains.
Traditionally, the notion of roles were defined based on graph equivalences such as structural, regular, and stochastic equivalences.
We briefly revisit these early notions and instead propose a more general formulation of roles based on the similarity of a feature representation (in contrast to the graph representation).
This leads us to propose a taxonomy of three general classes of techniques for discovering roles that includes (i) graph-based roles, (ii) feature-based roles, and (iii) hybrid roles.
We also propose a flexible framework for discovering roles using the notion of similarity on a feature-based representation.
The framework consists of two fundamental components: (a) role feature construction and (b) role assignment using the learned feature representation.
We discuss the different possibilities for discovering feature-based roles and the tradeoffs of the many techniques for computing them.
Finally, we discuss potential applications and future directions and challenges.
\end{abstract}

}

\maketitle
\IEEEdisplaynotcompsoctitleabstractindextext
\IEEEpeerreviewmaketitle

\section{Introduction}\label{sec:intro}
\PARstart{R}{ole discovery} first arose in sociology~\cite{parsons1951illness,merton1968social} where roles were used to explain the specific function of a person in society (such as a father, doctor, student, or an academic advisor)~\cite{borgatti2013book}.
These roles as defined by sociologists are known specifically as social roles~\cite{lorrain1971structural}.
From there, role discovery naturally became an important topic in social network analysis~\cite{hollandkathryn1983stochastic,arabie1978constructing,anderson1992building}.

In the past, role discovery has primarily been of interest to sociologists who studied roles of actors in extremely small offline social networks (e.g., graphs with tens of nodes)~\cite{anderson1992building,batagelj2004generalized,doreian2005generalized,nowicki2001estimation}.
Recently, role discovery is being explored in several other settings such as 
online social networks~\cite{scripps2007node},
technological networks~\cite{mahadevan2006internet,rossi2013topology}, 
biological networks~\cite{varki1993biological,luczkovich2003defining}, 
web graphs~\cite{ma2012mining},
among many others~\cite{golder2004social}.
While the concept of role discovery is indeed important for general graph mining and exploratory analysis, it can also be useful in many practical applications.
For example, roles might be used for detecting anomalies in technological networks such as IP-traces~\cite{mahadevan2006internet,rossi2013topology}.
An anomaly in this setting might be a node that doesn't fit any of the roles (normal structural patterns) or it may also be defined as a role that deviates from the common/normal roles, so that any node assigned to this unusual role would be anomalous~\cite{rossi2013dbmm-wsdm}.
Another use might be in online advertising campaigns~\cite{nk2013ads} for online social networks (Facebook, Groupon, Yelp) where ads could be customized based on the users' roles in the network. 
In addition, a business might only be interested in advertising to a person with a certain role in the network.
Furthermore, roles are becoming an important tool with potential applications such as 
classification, active learning, network sampling, anonymization, among many others (see Table~\ref{table:applications} for a summary).
Despite the various applications, the task of role discovery has only received a limited amount of attention (e.g., compared to the task of community partitioning~\cite{clauset2004finding,chen2012dense,Backstrom:2006,Chakrabarti:2006,yang2012spectral,newman2004fast}).

Role discovery was first defined as any process that divides the nodes of a graph into classes of structural equivalent nodes ~\cite{lorrain1971structural}.
These classes of structural equivalent nodes are known as roles.
Intuitively, two nodes are {\emph structurally equivalent} if they are connected to the rest of the network in identical ways.
There have been many attempts to relax the criterion of equivalence, e.g., regular equivalence~\cite{white1983graph}, stochastic equivalence \cite{holland1981exponential}.
For practical purposes, the notion of equivalence can be generally relaxed to get at some form of \emph{structural similarity}.
Thus, in this work, we replace the notion of equivalence by the notion of similarity (a weaker but more practical notion). 

\begin{figure}[t!]
    \centering
    	\hspace{-4mm}
     	\includegraphics[width=2.9in]{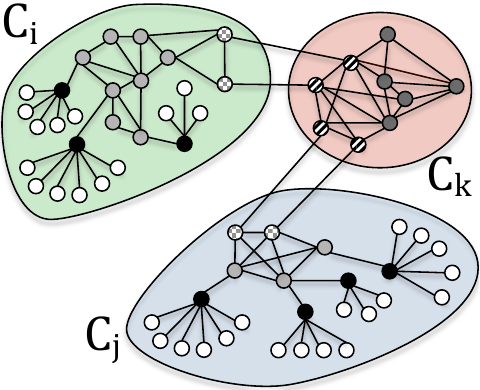}
\caption{
In the example above, the three communities (i.e., $C_i$, $C_j$, $C_k$) are cohesive sets of nodes. We also note that the nodes with similar shades/markings are assigned the same node-centric roles.
As shown above, roles represent structural patterns and two nodes belong to the same role if they have similar structural patterns.
Furthermore, notice that the nodes in a community are close together (i.e., distance/proximity-wise) whereas the nodes in a particular role are not required to be connected or close in terms of graph distance, they only need to be more structurally similar to that role than any other.
The example above is based on observed roles in large-scale technological networks such as the Internet AS topology.
In that example, nodes are assigned to roles based on the notion of regular equivalence~\ref{sec:reg-eqv}. As such, two nodes are ``regularly equivalent'' (play the same role) if they connect to equivalent others.
Clearly, nodes in the same role (e.g., star-center nodes) may be in different communities and thus the notion of role is independent of distance/proximity whereas nodes in the same community must be close to one another (small distance/proximity as defined by the fundamental property/notion of a community).
Another significant difference is that roles generalize as they may be learned on one graph and extracted over another, whereas communities do not.
}
\vspace{-3mm}
\label{fig:roles-and-comms-example}
\end{figure}

We can now redefine role discovery appropriately using this relaxation.
\textit{Role discovery}, informally speaking, is any process that takes a graph and picks out sets of nodes with similar structural patterns. 
Intuitively, two nodes belong to the same role if they are {\emph structurally similar}.
Roles of a graph represent the main node-level connectivity patterns such as star-center/edge nodes, peripheral nodes, near-clique nodes, bridge nodes that connect different regions of the graph, among many other types of connectivity patterns (See Figure~\ref{fig:roles-and-comms-example}).
In that example, the roles are defined in a strictly local sense, but in general roles may represent extremely complex node-centric structural patterns that fundamentally depend on the domain and underlying process that governs the graph.
Feature-based roles naturally generalize beyond the notion of structural similarity to the notion of similarity between features which includes pure structural features as well as features learned from an initial set of (non-relational) attributes (see Section~\ref{sec:foundations-feature-based-roles} and Section~\ref{sec:feature-construction} for more details).

As an aside, the notion of ``structural'' roles is significantly different than that of communities (and thus outside the scope of this survey).
More specifically, communities are sets of nodes with more connections inside the set than outside~\cite{clauset2004finding}, whereas roles are sets of nodes that are more structurally similar to nodes inside the set than outside.
Intuitively, roles represent ``significant'' structural patterns (e.g., star-centers, near-cliques, bridge-nodes, star-edge nodes) and two nodes belong to the same role if they have similar structural patterns~\cite{borgatti2013book,rossi2013dbmm-wsdm}.
In contrast, community methods assign nodes to sets based on the notions of density (and cohesion, proximity/closeness in the graph)~\cite{clauset2004finding}.
Therefore, it is clear that community methods have a completely different objective (partition nodes into sets based on density/cohesion/proximity) and often assigns every node to a community.
Figure~\ref{fig:roles-and-comms-example} illustrates a few of these differences and gives further examples.

\begin{table*}
\centering
\vspace{-3mm}
\caption{
\vspace{-2mm}
\textbf{Summary of the role tasks and applications}. 
We summarize a few of the tasks where the feature-based roles might be useful.
} 
\label{table:applications}
\def\arraystretch{1.3}
\scriptsize
\begin{tabular}{ r  p{146mm} }
\toprule
\multicolumn{1}{r}{\small \textsc{Role Tasks}}
& {\small \textsc{Goal/definition}}
\\ 
\bottomrule

\textbf{Dynamic roles} & \ZZ \XX
Use roles for descriptive and predictive modeling of dynamic networks\cite{rossi2013dbmm-wsdm,xing2010state}.
\\

\textbf{Classification} &
Use role memberships as features for statistical relational learning algorithms (relational/collective classification)~\cite{aha:09,neville:kdd03,mcgovern2008spatiotemporal}.
\\

\textbf{Visualizations} &
Roles may be used to visualize and capture the relevant differences and important patterns in big graph data~\cite{blei:jmlr08,xing2010state,rossi2012role-www}.
\\

\textbf{Graph similarity} &
Given two graphs $\gG$ and $\gH$, extract features and roles from each, and compare them~\cite{rossi2013topology}.
\\

\textbf{Entity resolution} &
Given two graphs $\gG_i$ and $\gG_j$ (e.g., Facebook and twitter social networks), use feature-based roles to resolve node entities (predict which entity/node in $\gG_i$ pertains to the exact same node in $\gG_j$)~\cite{bhattacharya:07}.
\\

\textbf{Anomaly detection} &
Given a graph $\gG$, find anomalous nodes (or links) with unusual role-memberships (static graph-based anomaly) or nodes with unusual role transitions (dynamic graph-based anomaly)~\cite{rossi2013dbmm-wsdm}.
\\

\textbf{Transfer learning} &
Learn roles on the graph $\gG_i$, and use these to learn the same set of roles on another network to improve accuracy~\cite{pan2010survey,rolX2012kdd}.
\\

\textbf{Graph compression} &
Roles capture the main structural patterns in a graph and may be used as a lossy graph compression technique~\cite{blei:jmlr08,rossi2012role-www}.
\\

\textbf{Queries/search} &
Find the top-k nodes (links) with the most similar roles (structural behavior)~\cite{rossi2012role-www}.
\\

\textbf{Active learning} &
Instead of selecting nodes via communities~\cite{bilgic2010active}, roles may be used to select nodes with diverse structural patterns/roles.
\\

\textbf{Anonymization} &
Use roles to compute a new representation to preserves the privacy and anonymize the vertex/edge identities~\cite{tassa2013anonymization}.
\\

\textbf{Network sampling} &
Sample a network based on the feature-based roles to ensure that all roles are represented in the sampled graph~\cite{nk2012network}.
\\

\bottomrule
\end{tabular}
\vspace{-5mm}
\end{table*}

\subsection{Scope of this Article}
This article focuses on examining and categorizing techniques for computing roles from a feature-based representation.
First, we propose a taxonomy for role-discovery which includes graph-based roles and feature-based roles.
After briefly reviewing the traditional graph-based roles in Section~\ref{sec:foundations}, the remainder of the article focuses on feature-based roles.
In particular, we propose the general problem of feature-based roles and provide a taxonomy and framework for it.
We formulate feature-based roles from a machine learning perspective which provides a natural basis for surveying and reinterpreting techniques for use in discovering roles.
Our feature-based role discovery framework succinctly characterizes the {\emph application-specific decisions} for discovering feature-based roles such as the types of features to use (graph-based, node, link or non-relational attributes), the graph feature operators to use (e.g., the specific aggregates, set ops, path/walk-based measures, subgraph patterns), whether to learn features automatically or manually, among many other decisions including role assignment methods for a feature-representation and selecting the number of roles (model selection).

Due to the no-free-lunch theorem~\cite{wolpert1997no,wolpert1996lack}, it is impossible for a single role discovery method to always work better than another, given that there is no assumption on the data used in learning roles~\cite{wolpert2002supervised,xu2012sparse,wolpert1997no,goutte1997note,wolpert1996lack}.
Therefore, we instead propose a general framework for feature-based roles that can serve as a fundamental basis for understanding, analyzing, and comparing methods for discovering feature-based roles.
In addition, the framework provides a guide for designing role discovery techniques that are suitable for user-defined applications with known phenomenon/assumptions.
Along these lines, we survey and discuss the relevant techniques, decisions, and their advantages and disadvantages, and interpretation for roles.
In addition, some techniques surveyed were used for other tasks and are reinterpreted/adapted for the role discovery problem (i.e., significant node-centric structural patterns).
This article does not attempt to survey types of blockmodels~\cite{goldenberg2010survey} or other types of graph-based roles, nor do we attempt to survey community partitioning and related methods as these focus on an entirely different goal/objective than the one examined in this article.

\subsection{Approach and Organization of Article}
This article proposes a taxonomy for the role discovery problem which includes both learning of node and edge roles using a graph-based representation and a feature-based representation.
Initially, roles have been computed using the graph representation directly -- i.e, graph-based roles (graph $\rightarrow$ roles), (e.g. stochastic block models \cite{nowicki:01,hollandkathryn1983stochastic,arabie1978constructing,anderson1992building,batagelj2004generalized,blei:jmlr08}). 
Recently, there has been a trend of research for computing roles from a feature representation -- i.e, feature-based roles (graph $\rightarrow$ features $\rightarrow$ roles). 
The feature representation process includes all the methods that can be used to transform the graph into an appropriate set of features for which roles can be defined over. 
More precisely, we transform the graph into a new representation (feature-based) from which the equivalences for the roles are computed. The set of features can be computed from node or link features, and non-relational attributes. 
While this paper aims to survey both graph-based and feature-based roles, we additionally provide the foundations of a generic framework for feature-based role discovery.
Note that one may also derive roles using a logical representation such as the work in ~\cite{mlnpedro2006,mlnrolelabeling}.

\medskip
\noindent This survey also makes the following contributions:

\begin{compactitem}[{\small $\bullet$}]
\setlength{\parskip}{3pt}
\item A precise formulation and taxonomy of the role discovery problem including motivation, techniques, evaluation, and applications
\item Brief survey of the current research in role discovery
\item A generic framework for feature-based role discovery
\item A taxonomy for constructing features for roles
\end{compactitem}
\medskip

This article is organized as follows.
Section~\ref{sec:foundations} discusses the difference between traditional roles and our feature-based roles.
This includes definitions and algorithms for computing each of these types of roles.
We also provide examples of roles and communities and discuss the differences of each task.
In Section~\ref{sec:framework}, we propose a general framework for discovering \textit{feature-based roles} and give an overview of it.
Section~\ref{sec:feature-construction} surveys and discusses past work on feature construction and reinterprets it for role discovery.
Afterwards, Section~\ref{sec:role-assignment} discusses techniques that can be used to assign roles to vertices based on the set of learned features.
Section~\ref{sec:challenges} discusses challenges and new directions.

\section{Role Foundations} \label{sec:foundations}
Role discovery can be generally defined as any process that divides the nodes into classes of equivalent nodes.
These classes of equivalent nodes form the foundation of what are known as roles.
Note that the classes here can be thought of social functions for the nodes in the network (e.g., a function may be a father or student).
However, the above definition relies on some formal notion of node equivalence which is used to divide the nodes into their respective roles.  
As such, all nodes that have a certain role should be equivalent under the predefined node equivalence relation. 
More precisely, assume we are given a graph $G=(V,E)$, let $r(u)$ and $r(v)$ be the role class of nodes $u$ and $v$ respectively -- $\forall u,v \in V, {r(u)=r(v)} \iff {u \equiv v}$. 
Now, the question is how to define the node equivalence relation. 
For example, are two nodes equivalent if they have connections to exactly the same neighbors?

\begin{figure}[h]
\centering
\hspace{-5mm}
\includegraphics[width=1.02\linewidth]{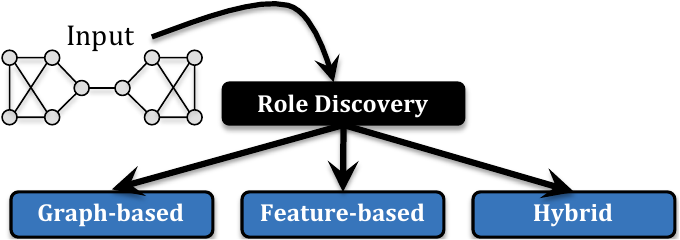}
\caption{
\textbf{Taxonomy of Role Discovery Methods}. 
We categorize methods for computing roles into graph-based roles, feature-based roles, and hybrid approaches.
Graph-based roles are computed from the graph directly (without incorporating any features).
In contrast, feature-based roles are computed from a transformation of the graph data into a new feature representation for which roles are discovered.
Naturally, there can also be hybrid approaches that leverage the advantages of both.
}
\vspace{-4mm}
\label{fig:role-taxonomy}
\end{figure}

We briefly review the fundamental node equivalences as they will be useful to analyze the different types of roles (graph-based, feature-based, and hybrid role methods) and the corresponding approaches for each type.
See Figure~\ref{fig:role-taxonomy} for an intuitive taxonomy of the three fundamental types of role discovery methods.

\subsection{Graph-based Role Equivalences}\label{sec:role-notions}
We first review the important equivalences that have formed the basis for roles in much of the literature. 
The equivalences are discussed below from most strict to least.
Let us note that one possible direction for future work will be the extension or formulation of these equivalences for role discovery in temporal or streaming networks~\cite{fu2009dynamic,xing2010state,rossi2013dbmm-wsdm,rossi2012role-www}.

\subsubsection{Structural equivalence}\label{sec:structural-equivalence}
Structural equivalence~\cite{lorrain1971structural} states that equivalent nodes have the same connection pattern to the exact same neighbors (i.e., in/out neighbors if edges are directed).
Therefore, two nodes $v_i$ and $v_j$ are structurally equivalent if they have the same exact neighbors, hence $\N(v_i) = \N(v_j)$.
For example, the set $W$ of nodes in Figure~\ref{fig:equiv} are structurally equivalent since each node connects to exactly the same (identical) neighbors.
This implies that structurally equivalent nodes are essentially indistinguishable in the sense that they have same degree, clustering coefficient, centrality, belong to the same cliques, etc.
Obviously, structural equivalence is too strict of a notion and thus impractical for (large) real-world graphs.
The fundamental disadvantage of structural equivalence is that it confuses similarity with closeness.
This arises due to the requirement that two nodes have the same \emph{exact} neighbors.
In fact, nodes that are structurally equivalent can never be more than two links away.
As a result, there have been many relaxations of structural equivalence~\cite{everett1985role,sailer1979structural,everett1990ego,holland1981exponential,white1983graph}.

\subsubsection{Automorphic equivalence}
A mapping $p$ from one graph to another is an isomorphism if whenever $u \rightarrow v$, then $p(u) \rightarrow p(v)$.
Isomorphisms are mappings from one graph to another that preserve the structure of a graph.
An automorphism is an isomorphism from one graph to the \emph{same} graph, thus preserving the symmetries.
A node $u$ is \emph{automorphically} equivalent to node $v$ if there exists an automorphism $p$ such that $u = p(v)$ ~\cite{holland1981exponential}.
As an aside, automorphic equivalence can be viewed as a relaxation of structural equivalence since any set of structural equivalences are also automorphic equivalences.
More intuitively, structural equivalence essentially asks if a \emph{single node} can be exchanged for another while preserving the connections/relationships of that node, whereas automorphic equivalence is based on \emph{sets of nodes} whom are exchangeable as subgraphs.
In Figure~\ref{fig:equiv}, the nodes in the combined set $W \cup M$ form a single class of exchangeable nodes (also known as a role).
Similarly, the nodes $v_1$ and $v_2$ from Figure~\ref{fig:equiv} form another role since both nodes are exchangeable if nodes of other classes are also exchanged.

\subsubsection{Regular equivalence}
\label{sec:reg-eqv}
\begin{figure}[t!]
\centering
\hspace{-4mm} \includegraphics[width=2.8in]{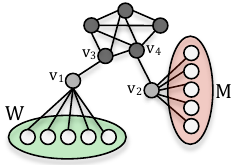}
\vspace{-2mm}
\caption{
This illustration reveals the fundamental differences between the main role equivalences including structural, automorphic, regular, and stochastic equivalences.
For example, the set of nodes $W$ are structurally equivalent forming a role whereas the nodes in $M$ are structurally equivalent with one another forming another distinct role, whereas the combined set of nodes $W \cup M$ are automorphically equivalent thus representing a single role when using the relaxed automorphic equivalence.
}
\vspace{-2mm}
\label{fig:equiv}
\end{figure}

Regular equivalence relaxes the notion of role further to capture the social role concept better.
In particular, regular equivalence is based on the idea that nodes play the same role if they are connected to \textit{role-equivalent nodes}. 
This is in contrast to structural equivalence where the nodes have to be connected to identical nodes.
In other words, regular equivalence states that nodes play the same role if they have similar connections to nodes of other roles\footnote{Nodes that are structurally or automorphically equivalent are also regularly equivalent, but the inverse is not true.}~\cite{sailer1979structural,white1983graph}.
Intuitively, two nodes that are regularly equivalent (i.e., same role) do not necessarily have to connect to the same neighbors or even the same number of neighbors, but they must be connected to \emph{role-equivalent neighbors}.
From Figure~\ref{fig:equiv}, notice that the nodes in the set $W \cup M$ are regularly equivalent (unique role), while $v_1$ and $v_2$ form another class of regularly equivalent nodes (2nd role).
Further, $v_3$ and $v_4$ represent the 3rd role while the last three nodes form the fourth class of regularly equivalent nodes.
For instance, the nodes in the third role are $v_3$ and $v_4$, since all nodes in that set have at least one edge to a node in the 2nd role, while also connected to nodes in the fourth role.
Notice that regular equivalences can be exact or approximate, and hence there might be many valid ways of grouping nodes into equivalence sets for a given graph.
However, this definition is still strict in the sense that the role of a node is tied to all nodes of that role, rather than some nodes~\cite{boyd1999relations}.

\subsubsection{Stochastic equivalence}
For this reason, stochastic equivalence was introduced~\cite{holland1981exponential}, which informally says that for a probability distribution of edges in a graph, an assignment of roles is a stochastic equivalent, if nodes with the same role have the same probability distribution of edges with other nodes.
Intuitively, the nodes are organized into roles such that the the probability of a node linking with all other nodes of the graph are the same for nodes of the same role. 
Another interpretation is that the probability distribution of the graph must remain the same when equivalent nodes are exchanged~\cite{nowicki:01}.
From the example in Figure~\ref{fig:equiv}, observe that stochastic equivalence gives rise to the same equivalence classes obtained using the notion of regular equivalence due to the simplicity of the graph. 
This notion gives rise to stochastic blockmodels~\cite{nowicki:01,hollandkathryn1983stochastic,arabie1978constructing,anderson1992building,batagelj2004generalized}, which comes from weakening and extending the algebraic notion of structural equivalent nodes.

\subsection{Methods for Graph-based Roles}\label{sec:graph-based-roles}
We can define graph-based roles as those computed directly from the graph representation, which is typically in the form of an adjacency matrix.
This is in contrast to feature-based roles which are computed indirectly from the graph by first transforming the graph representation into a feature-vector representation.
Feature-based roles are the main focus of this paper, but for completeness we briefly discuss a few approaches for computing graph-based roles.

\subsubsection{Blockmodels}\label{sec:blockmodels}
Blockmodels are by far the most popular class of role techniques in social network analysis~\cite{hollandkathryn1983stochastic,arabie1978constructing,nowicki:01,doreian2004generalized,blei:jmlr08,fu2009dynamic,xing2010state,doreian2005generalized}. 
Blockmodels naturally represent a network by a compact representation known as a role-interaction graph (or image matrix) where the nodes represent roles (or blocks, positions) and the edges are interactions between roles.
This smaller comprehensible structure can be interpreted more readily. 
These methods attempt to group nodes according to the extent that they are structurally equivalent or stochastically equivalent. 
Recently, many types of blockmodels have been proposed such as stochastic blockmodels~\cite{nowicki2001estimation}, generalized blockmodels~\cite{batagelj2004generalized}, and mixed-membership stochastic blockmodels~\cite{blei:jmlr08} (MMSB).
They have also been extended for various applications~\cite{doreian2004generalized,blei:jmlr08,fu2009dynamic,xing2010state}.
Nevertheless, a complete survey of these graph-based role methods are beyond the scope of this paper, see~\cite{goldenberg2010survey} for more details.
However, we do provide a brief overview of some of the main methods below.

One of the first methods used for computing a type of blockmodel was CONCOR (convergence of iterated correlations), which was proposed by Breiger \etal~\cite{breiger1975algorithm}.
This method initially computes the correlation of the adjacency matrix $\mathsf{Corr}(\mA,\mA)$ which we denote as $\mC_0$, then the correlation matrix $C_1$ is computed from the previous correlation matrix $\mC_0$, and this process is repeated until all entries are either $1$ or $-1$.
Let us note that even though this method was originally designed for blockmodels, it is fundamentally based on similarity using the adjacency matrix and does not have an explicit criterion function in the usual sense of an optimization problem.
On the other hand, blockmodels are typically formulated as optimization problems with a well-formed objective function.
However, in general, there are many ways to compute blockmodels (e.g., direct and indirect approaches)~\cite{batagelj1992direct,batagelj2004generalized,batagelj2002generalized,batagelj2007indirect}.

Stochastic blockmodels (SBM)~\cite{nowicki:01} on the other hand adopts the notion of stochastic equivalence~\cite{holland1981exponential}.
The benefit is that these probabilistic models allow for deviations between the observations and relaxes the idealized concept of exact equivalence.
These models are a type of latent space models~\cite{lazarsfeld1968latent}, but also fall into the category of random graph models~\cite{goldenberg2010survey}.
One simple example of a blockmodel is a model that assigns each node to one of the several roles (or blocks)\footnote{The number of roles $k$ can be specified by the user or learned from the data.}.
In addition, there is a set of probabilities $p_{i,j}$ that specifies the probability of an edge between a node in role $i$ and a node in role $j$.
In other words, how likely is it that a node in role $i$ has an edge between a node in role $j$?
This is typically represented as a matrix $\mP \in \mathbb{R}^{k \times k}$ and can be either specified by the user or inferred from the data.
In general, these models can be used to generate a random graph by specifying the probabilities in $\mP$ or these probabilities may be inferred from data.
Essentially, MMSB allows nodes to take part in multiple roles while also allowing roles to link to other roles probabilistically.), then one could place large weights on the diagonal of $\mP$, which indicates that nodes of the same role have a large probability of having an edge between each other and low probability of having an edge to a node in another role.

More recently, Airoldi \etal~\cite{blei:jmlr08} proposes a mixed-membership stochastic blockmodel (MMSB) that relaxes the assumption of nodes belonging to only a single role. 
The model is instantiated using a blockmodel and combines this with mixed-membership~\cite{erosheva2005bayesian}.
We also note that there are other various models such as the latent space models~\cite{hoff2002latent}.

\subsubsection{Row/Column Similarity of Adjacency Matrix} \label{sec:sim-adj}
While the blockmodels are the most popular, there are also other techniques for computing graph-based roles that use some form of similarity between the rows of the adjacency matrix~\cite{burt1976positions}.
These graph-based similarity role methods have two general steps:
First, the similarity (or distance) is computed between each pair of rows in the adjacency matrix.
For this, any traditional similarity measure such as euclidean distance or correlation can be used.
After computing the similarity matrix, the second step clusters the nodes using this similarity matrix.
For the clustering, any traditional method can be used (e.g., hierarchical clustering and multi-dimensional scaling (MDS)~\cite{kruskal1964nonmetric} are the most common).

There are also spectral methods that compute the eigenvectors of the adjacency matrix (or a similarity matrix) then uses some subset of these to derive roles~\cite{brandes2010structural}.
For computing roles (structural patterns), the interesting eigenvectors are not only those with the largest eigenvalues, but rather a subset of the eigenvectors that represent distinct structural patterns (stars-centers, star-edges, bridges, near-cliques, etc.).
This arises due to the fact that role discovery methods attempt to capture or model all the significant structural patterns present in the data and are not restricted or focused on only a single type of pattern.
As an aside, the eigenvectors associated to the largest eigenvalues represent one type of role (i.e., usually near-clique or tightly connected nodes).

Nevertheless, there has been some work focused on finding certain ``types'' of nodes in the graph that have a particular predefined role (i.e., structural pattern)~\cite{kleinberg1999hubs,tong2012gateway,jiang2009mining}.
For example, an early work by Kleinberg \etal~\cite{kleinberg1999hubs} essentially computes the Singular Value Decomposition (SVD)~\cite{golub1970singular} of a graph's adjacency matrix (i.e., eigenvectors of $\mA\mA^T$ and $\mA^T\mA$)\footnote{$\mA\mA^T$ and $\mA^T\mA$ are similarity matrices.}, then identifies two types of star-center nodes based on incoming or outgoing edges known in that work as authority nodes and hub nodes~\cite{chung1997spectralbook}. 
These nodes were of interest as they relate to pages on the web that serve as a portal (hub) with many outgoing edges to reputable sites and pages that have many incoming edges (authorities).

\subsubsection{Discussion}
The main disadvantage of the models in Section~\ref{sec:blockmodels} is that they are difficult to compute for large graphs and even more so for the massive graphs found in the real world such as Facebook's friendship graph, Twitter who-tweets-whom, among many others.
Most of the previously mentioned work evaluates their model using rather small networks.
For example, one recent model, dMMSB, which extends MMSB for dynamic networks, takes around a day to compute for 1,000 nodes, since their method is quadratic in the number of nodes.

This is in contrast to techniques based on similarity of the adjacency matrix (in Section~\ref{sec:sim-adj}), which if properly configured can be fast to compute~\cite{brandes2010structural} .
The disadvantage of these methods is that the roles may be less meaningful or harder to interpret, whereas those based on blockmodels may be more accurate or easier to interpret from the implicit role definition implied by the model. 
Nevertheless, the utility of the roles ultimately depends on the application (what they are used for) and/or domain (where they are used).

\subsection{Feature-based Roles} \label{sec:foundations-feature-based-roles}
We first define feature-based roles and propose a taxonomy to illustrate the difference between graph-based roles.
In Section~\ref{sec:feature-node-equivalence} we discuss some possibilities of extending node equivalences for feature-based roles, then Section~\ref{sec:feature-role-methods} discusses a few of the past approaches, and finally end with a discussion.

Intuitively, feature-based roles are derived by transforming the graph representation into a feature representation, then assigning roles based on some notion of feature equivalence.
This is in contrast to the previous approaches in Section~\ref{sec:graph-based-roles} that compute roles directly from the graph representation~\cite{goldenberg2010survey,nowicki:01,blei:jmlr08}.
Roles computed from a feature representation are denoted as feature-based roles.
The set of features typically arises based on some transformation(s) over the graph, $f(\mathcal{G}) = \mX$ where $\mathcal{G}$ is a graph, $\mX$ is the set of features, and $f(\cdot)$ is some collection of transformations over the graph. 
A transformation may be a set of aggregates or any other feature operator, see Table~\ref{table:feature-ops}.
More generally, the set of features may arise through some transformation(s) over an attributed-graph such that  $f(\mathcal{G}$, $\mX) = \tilde{\mX}$ where the input features $\mX$ and output $\tilde{\mX}$ may consist of node features, link features, or non-relational features.

\begin{figure}[b!]
\vspace{-5mm}
\centering
\includegraphics[width=1.8in]{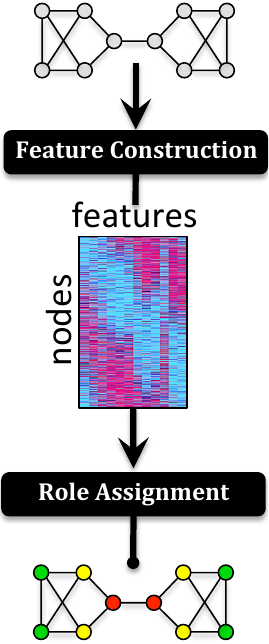}
\caption{\textbf{General Framework for Feature-based Roles.}
The framework consists of \emph{role feature construction} and \emph{role assignment}.
}
\label{fig:feature-based-roles-process}
\end{figure}

\subsubsection{Node Equivalence for Feature-based Roles}\label{sec:feature-node-equivalence}
The work on graph-based roles adopts some notion of node equivalence with respect to the the graph representation~\cite{lorrain1971structural,white1983graph,everett1994regular,holland1981exponential,hollandkathryn1983stochastic}.
Since feature-based roles are computed from a different representation, we must move from the notions of node equivalence on the graph representation to node equivalence on a \emph{feature representation}.
We view node equivalence on a feature-based representation in two main ways.
First, one may develop feature-based role methods that are consistent with one of the previous node equivalences (e.g., regular equivalence).
Second, we may simply reinterpret/extend these equivalences for computing roles from a feature representation.
Besides extending these definitions for features, one may also relax the node equivalences as done in graph-based roles.
In this work, we primarily focus on the second view as this allows for greater flexibility, efficient methods, and may be more useful from a practical point of view.
\begin{Definition}[Node Equivalence on Feature Representation]
Let $f_1, f_2, ..., f_m$ be a collection of structural features (degree, distance, ...), and let $u$ and $v$ be two arbitrary nodes, then a strict notion of node equivalence is defined as,
\[ (\forall i, 1 \leq i \leq m: f_i(u) = f_i(v)) \Rightarrow u \equiv v \]
\end{Definition}
This is strict in the sense that two nodes share the same role if they have identical feature-vectors.
However, a common trend in the past has been to relax the notions of node equivalences while maintaining the properties of interest~\cite{holland1981exponential,white1983graph,nowicki:01,blei:jmlr08}.
For instance, most previous work is based on relaxations of structural equivalence, e.g., regular equivalence, stochastic equivalence, automorphic equivalence, among others.

\subsubsection{Node Similarity for Feature-based Roles}
As such, for feature-based roles, we can also move from the strict notion of node equivalence to the relaxed notion of node similarity (on a feature representation).
Intuitively, two nodes $u$ and $v$ share the same role, if they have similar feature-values. 
\begin{Definition}[Feature Similarity (Equivalence)]
Two nodes $u$ and $v$ are similar, if $\mathsf{S}(\vx_{u}, \vx_{v}) \approx 1$ where $\mathsf{S(\cdot)}$ is a standard similarity (or distance measure: $\mathsf{D(\cdot)} = 0$), and $\vx_{u}$ and $\vx_{v}$ are the feature-vectors of $u$ and $v$, respectively.
The similarity function $\mathsf{S}(\cdot) = 1$ if $\vx_u$ and $\vx_v$ are identical (s.t. 1 is the maximum similarity).
\end{Definition}
Nevertheless, if the features are strictly graph features and representative of the structural properties in $\mathcal{G}$, then this implies $u$ and $v$ are structurally similar.
We note that methods for computing a low-rank approximation or clustering (hierarchical) typically use some form of similarity (e.g., euclidean distance, Frobenius norm, ...).
Notice that the above definition is independent of the neighbors of nodes and relies only on the features.
This independence avoids roles being tied to each other based on cohesion (as is a problem of structural equivalence, see Section~\ref{sec:structural-equivalence}).
Further, the definition is based on some notion of similarity, which is typically a relaxation of the stricter notion of structural equivalence which we can easily reinterpret for features.
In terms of flexibility, these models are capable of expressing a larger class of roles than graph-based methods.
This is a result of representing roles from a set of features constructed from a possibly infinite feature space.

\subsubsection{Inspiration \& Examples} \label{sec:feature-role-methods}
Feature-based roles were perhaps inspired by indirect approaches to blockmodeling~\cite{batagelj2002generalized,batagelj2007indirect,batagelj1992direct}, latent feature models which are more expressive than traditional blockmodels~\cite{miller2009nonparametric,griffiths2005infinite,navarro2008latent,ghahramani2007bayesian,doshi2009correlated}, and other approaches based on the similarity of the rows of the adjacency matrix~\cite{burt1976positions,brandes2010structural}.
We briefly review a few approaches for computing feature-based roles.

The first such method for computing a type of feature-based role was initially explored by Batagelj \etal \cite{batagelj1992direct}.
In this seminal work, they formally define what it means for a collection of structural features (degree, distance, ...) to be consistent with structural equivalence~\cite{batagelj1989similarity}.
This indirect approach measures the equivalence of nodes based on the feature-values.
Let us note that Burt \etal~\cite{burt1983applied} also proposed a dissimilarity measure consistent with structural equivalence.
These indirect approaches were originally used in block modeling problems~\cite{batagelj2002generalized,batagelj2007indirect,batagelj1992direct} and were usually designed to be consistent with one of the previous node equivalences from Section~\ref{sec:role-notions}.

Some other work has recently focused on a more general type of feature-based role~\cite{rossi2012role-www,rolX2012kdd,rossi2013dbmm-wsdm}, which are not consistent with respect to the previous node equivalences defined by sociologists.
In general, these approaches use some form of structural similarity for features.
Intuitively, two nodes share the same role, if they have similar features.
These approaches construct a large set of local features, that are specially tuned for social networks, then uses Non-negative Matrix Factorization (NMF)~\cite{lee1999learning} to assign roles to the nodes.
Some work has even used this particular specification of feature-based roles (i.e., degree/egonet features, NMF, Akaike's information criterion (AIC)~\cite{akaike1974new}) to model and explore dynamic networks~\cite{dbmm11-tr} and more recently to improve the accuracy of classification~\cite{rolX2012kdd}.
Let us note that this work only explores one type of feature-based role, while there exists many other opportunities that are perhaps more accurate, faster, or require tuning for specific applications.
In Section~\ref{sec:framework}, we propose a general framework for discovering feature-based roles, and discuss the issues, decisions required, and more generally the opportunities for using feature-based roles. 

\subsubsection{Discussion}
Traditional graph-based approaches may not be flexible enough to capture complex roles in large networks due to their limitations in expressing such roles.
For instance, McDaid~\cite{mcdaid2012model} identifies two roles for the small karate club network, where the first role corresponds to high degree nodes and the second corresponds to small degree nodes.
Alternatively, this article introduces a general framework for computing roles from a feature-based representation.
We argue that feature-based roles provide greater flexibility for representing complex roles that are frequently seen in the real world and likely to be important for practical applications.
On the other hand, the flexibility provided from a feature-based representation makes it more difficult to identify the features required to capture the roles warranted by the researcher.
This part of the process takes some guidance by the expert or more tuning for a specific application.

\subsection{Hybrid approaches}\label{sec:hybrid-roles}
At the intersection of graph (Section~\ref{sec:graph-based-roles}) and feature-based roles (Section~\ref{sec:foundations-feature-based-roles}) lies hybrid approaches that leverage both the graph and feature representation (typically learned from a relational learning system~\cite{davis2005integrated,landwehr2005nfoil,neville:kdd03,rossi2012drc,mcgovern2008spatiotemporal}) in some fashion.
We categorize hybrid role discovery methods into two main classes based on if a graph-based approach is used prior to role feature construction or whether the graph structure is leveraged after the learning of the feature representation for roles.

The first class of hybrid role discovery methods use a graph-based approach prior to role feature construction.
For instance, we may use a blockmodel to extract roles directly from the graph, which can then be viewed as ``initial attributes'' for learning more sophisticated or targeted features.
Once the initial attributes are learned, we can give them along with the graph as input to a relational feature learning system for which more meaningful features can be learned by incorporating the knowledge from the initial graph-based role discovery method.
Using this refined set of features, we can now use a technique to automatically learn the number and assignment of the ``hybrid-based'' roles (See Section~\ref{sec:role-assignment}).
This approach can be seen as a way to loosely constrain the feature-based roles to be similar to the specific parametric form assumed by the statistical blockmodel.
The main disadvantage of these methods lies in their scalability due to the limitations of blockmodels (e.g., SBM~\cite{nowicki:01}, MMSB~\cite{blei:jmlr08}), but remain a promising direction in the future as these methods become more scalable (See Section~\ref{sec:challenges}).
Nevertheless, other more scalable graph-based approaches remain promising for this class of hybrid role discovery methods~\cite{kleinberg1999hubs}.

The second class of hybrid roles leverage multiple data sources as a way to regularize or influence the role assignment phase.
It is assumed that a feature representation $\mX$ from the graph was first learned, but there are additional data sources (i.e., graphs and attribute sets) that may be useful for role discovery.
These hybrid roles can be learned by adapting tensor factorization methods~\cite{de2009survey,friedlander2008computing,yilmaz2010probabilistic,cichocki2008advances,kim2007nonnegative} or collective matrix-tensor factorization (CMTF) methods~\cite{ma2008sorec,bouchard2013convex,singh2008relational}.
Unlike tensor factorization methods, collective factorization methods can learn roles by fusing multiple heterogeneous data sources represented as matrices and/or tensors.
For instance, we might have a store-categories matrix, a user-store-item tensor, and a social network matrix.
In that context, one might posit that the roles of the users should be influenced by the store as well as the items that were purchased from that store and its category.
Note these techniques may also consider time giving rise to dynamic roles (See Section~\ref{sec:dynamic-roles}).
More generally, collective matrix-tensor factorization allows for any number of matrices or tensors to be included in the factorization which in turn allows these to directly influence the learned role definitions.

\section{Framework for Feature-based Roles}\label{sec:framework}
This section introduces a flexible framework for \textit{feature-based role discovery} which offers many benefits over traditional graph-based roles by computing roles from a feature representation (instead of directly from the graph, see Section~\ref{sec:foundations}). 
In particular, the framework consists of the following basic computations:

\medskip
\noindent
\textbf{Role feature construction.} Transform graph into a set of graph features 
\medskip
\newline
\textbf{Role assignment.} Assign nodes with similar feature vectors to same roles
\medskip

The general framework, along with the two fundamental steps for discovering feature-based roles are intuitively illustrated in Figure~\ref{fig:feature-based-roles-process}.
In that illustration,
the graph is first transformed into a feature-based representation through some type of feature-construction technique or SRL system (section~\ref{sec:feature-construction}). 
Afterwards, roles are extracted from the large set of features via some type of low-rank approximation (matrix factorization) or clustering algorithm (section~\ref{sec:role-assignment}).

A detailed overview of the framework is shown in Table~\ref{table:framework}.
Importantly, the feature learning systems used for \textit{role feature construction} are interchangeable.
For instance, one may use a simple feature learning system that searches over the space of local 1-hop neighborhood features or one may use an entirely different system that searches over more global features.
The choice is entirely dependent on the important structural patterns present in the data and the overall application constraints such as scalability. 
Likewise, the methods for  \textit{role assignment} are also interchangeable as one might choose to use NMF instead of SVD for learning the roles from the set of learned features.
This allows the framework to be useful and flexible for application-specific role discovery.
Therefore, Table~\ref{table:framework} also highlights the main categories of techniques for both feature construction and role assignment.

A key aspect of the feature-based roles is in their flexibility.
For instance, unlike graph-based roles which use the graph directly, the feature-based roles (automatically) transform the graph (and any initial attributes) into a feature representation for which the role definitions are learned and extracted.
Moreover, the proposed framework can also be used for the novel task of computing roles on the edges of a graph. 
We also note that application-specific constraints such as sparseness, diversity, locality, among many others may be placed on the feature-based role discovery problem, in either the feature construction or role assignment.
The advantages of our general feature-based role framework are summarized below.
\smallskip
\begin{compactenum}[{\scriptsize $\bullet$\leftmargin=0em}]
\setlength{\parskip}{4pt}
\item Flexible framework for feature-based roles.
\item Roles can be easily tuned for specific applications. 
\item Complexity and efficiency may be balanced depending on application-specific constraints.
\item Able to capture arbitrary structural patterns (i.e., using data-driven and non-parametric approaches).
\item Roles generalize since they are defined over features.
\item Attributes are easily incorporated, since roles are naturally based on features they can simply be included as additional features before (or after) learning and constructing novel features automatically (see Section~\ref{sec:feature-construction}).
\end{compactenum}

\begin{table}[b!]
\vspace{-4mm}
\caption{
\vspace{-2mm}
\textbf{Overview of the Feature-based Role Framework.}
}
\vspace{-5mm}
\begin{center}
\footnotesize
\begin{tabular}{| c  c | l | } 
\noalign{\hrule height 0.5mm}

\TTZ \BBZ 
\multirow{6}[0]*{\rotatebox{90}{\textbf{\normalsize  Feature-based Roles}}}
& \multirow{3}[0]*{\textbf{\normalsize Role Feature}}
& \textbf{Section~\ref{sec:feature-classes}) Relational Feature Classes} 
\\  \cline{3-3}

\TTZ \BBZ
& \multirow{2}[0]*{\textbf{\normalsize Construction}}
& \textbf{Section~\ref{sec:feature-ops}) Relational Feature Operators} 
\\  \cline{3-3}

\TTZ \BBZ 
& \multirow{2}[0]*{\textbf{\large Section~\ref{sec:feature-construction}}}
& \textbf{Section~\ref{sec:feature-search}) Feature Search} 
\\  \cline{3-3}

\TTZ \BBZ  
& 
& \textbf{Section~\ref{sec:feature-selection}) Feature Selection} 
\\  \cline{3-3}
\Cline{10em}{2-3}
\Cline{10em}{2-3} \Cline{10em}{2-3} \Cline{10em}{2-3}
\Cline{10em}{2-3} \Cline{10em}{2-3}
\Cline{10em}{2-3} \Cline{10em}{2-3}

\TTZ \BBZ 
& \multirow{3}[0]*{\textbf{\normalsize Role Assignment}}
& \textbf{Section~\ref{sec:role-clustering}) Feature Grouping/Clustering}
\\  \cline{3-3}

\TTZ \BBZ
& \multirow{3}[0]*{\textbf{\large Section~\ref{sec:role-assignment}}}
& \textbf{Section~\ref{sec:low-rank-approx}) Low-rank Approximation}
\\  \cline{3-3}

\TTZ \BBZ 
& 
& \textbf{Section~\ref{sec:model-selection}) Model Selection}
\\  \cline{3-3}

\noalign{\hrule height 0.5mm}
\toprule
\end{tabular}
\end{center}
\vspace{-4mm}
\label{table:framework}
\end{table}

\smallskip
\noindent
We also note that at the data-level, there are a number of factors that influence the learning and tuning of feature-based roles.
Researchers and engineers may design a feature-based role method that is consistent in terms of the knowledge and a priori assumptions.
Such a feature-based role method is likely to be much more effective for the specific application at hand.
Assumptions and knowledge that may help guide the role discovery process include data and structure and/or future dependencies (e.g., sparsity, size, complexity, patterns), along with application/domain knowledge.

\section{Role Feature Construction}\label{sec:feature-construction}
In this section, we discuss the process for discovering features for roles.
Relational feature construction is the systematic generation of features based on the graph structure or non-relational information.
Roles have traditionally been defined strictly from the graph (e.g., graph-based roles from Section~\ref{sec:foundations}).
In this work, we make only basic assumptions on the input used to discover roles.
We assume that we are given a graph $\gG$ which may have some initial 
node attributes denoted $\mX^v$\footnote{$\fX$ is used for $\mX^v$ when meaning is clear} or 
edge attributes $\mX^{e}$.
For example, a node attribute might be gender (m/f) and a link attribute might be the text of an email sent between two users.
Throughout the process of feature construction, additional features may be added to set of features $\mX$.

The goal of feature construction from its traditional view in machine learning is to construct features that are highly correlated with the prediction variable while being uncorrelated with each other.
Unfortunately, the goal of feature construction for roles is not as straightforward.
Previously, roles were defined mathematically based on structural equivalences~\cite{lorrain1971structural}, which are far too strict for practical purposes.
Consequently, these notions were relaxed to get at the notion of structural similarity instead of equivalence~\cite{everett1994regular,holland1981exponential,hollandkathryn1983stochastic,blei:jmlr08}.
In this work, we use this same idea, but adopt it for computing roles from features, instead of the graph directly.
Informally, the goal of a feature construction system should be to generate a set of features that capture the fundamental structures and important patterns in the graph data.
We can further relax this definition by allowing the features to capture the interest to a specific domain (technological networks, social networks) or for a specific application (anomaly detection in computer networks).
Since this definition depends intrinsically on the domain and application, we focus on surveying and discussing general ways to construct features for role discovery.

Intuitively, there are four main steps for learning a feature representation for role discovery: 

\smallskip
\begin{compactenum}[{\small $-$} \leftmargin=0em]
\item \textbf{Relational Feature Classes (Section~\ref{sec:feature-classes}).} Select the types of features to construct based on the graph data used in the computation.
\item \textbf{Relational Feature Operators (Section~\ref{sec:feature-ops}).} Determine the operators to use to construct those types of features.
\item \textbf{Feature Search Strategy (Section~\ref{sec:feature-search}).} Select a strategy for searching over the feature space including exhaustive, randomized, and guided search methods.
\item \textbf{Relational Feature Selection (Section~\ref{sec:feature-selection}).} Determine how features are evaluated/scored and pruned (incrementally) during the learning process.
\end{compactenum}
\medskip

\noindent
See Table~\ref{table:feature-construction-taxonomy} for an overview.
We also present in Algorithm~\ref{alg:feature-construction-template} a generic algorithm for discovering features.

\begin{table}[t]
\vspace{-2mm}
\caption{
\vspace{-2mm}
\textbf{Taxonomy for Role Feature Construction}.
The proposed taxonomy for role feature construction is simple and intuitive, consisting of only the four main steps below.
These steps can be viewed as explicit decisions that need to be learned automatically via the data and machine learning techniques or customized manually by a user (usually for a specific application/task).
Note that this taxonomy expresses a very large family of algorithms for which a set of features can be computed for the role discovery problem.
}
\label{table:feature-construction-taxonomy}
\vspace{-2mm}
\begin{center}
\begin{tabular*}{0.95\linewidth}{|p{0.90\linewidth}|}
\hline
\T \B  \textbf{\normalsize Role Feature Construction Steps \&}
\normalsize \textbf{Examples}
\\ 
\hline
\T \B
\T \B \multirow{1}{*}{\rotatebox{0}{\textbf{\normalsize Section~\ref{sec:feature-classes}) Relational Feature Classes}}} 
 \begin{compactenum}[{\scriptsize$\bullet$}\leftmargin=0em]
 \item Graph features ($V,E$)~\cite{rossi2012role-www,rahman2012graft}
 \item Relational link-value features	($\X^e, V, E$)~\cite{rossi:pakdd12} 
 \item Relational node-value features ($\X^v, V, E$)~\cite{neville:kdd03} 
\item Non-relational features ($\X$)~\cite{breslow1997simplifying,pan2010survey}
 \end{compactenum}\\ \hline

\T \B \multirow{1}{*}{\rotatebox{0}{\textbf{\normalsize Section~\ref{sec:feature-ops}) Relational Feature Operators}}} 
\begin{compactenum}[\scriptsize$\bullet$\leftmargin=0em]
 \item Aggregates (mode, mean, count, ...)~\cite{neville:srl00,rossi2013dbmm-wsdm}
 \item Set ops (union, intersection, multiset)~\cite{kohavi1997wrappers,aha:09}
 \item Subgraph patterns (2-star, 3-star, triangle, etc.)~\cite{robins:sn07,rahman2012graft}
 \end{compactenum}
\\
\hline
\T \B \multirow{1}{*}{\rotatebox{0}{\normalsize \textbf{Section~\ref{sec:feature-search})} \textbf{Feature Search Strategy}}} 
\begin{compactenum}[\scriptsize$\bullet$\leftmargin=0em]
 \item Exhaustive~\cite{neville:kdd03} 
 \item Random~\cite{mcgovern2008spatiotemporal}
 \item Guided~\cite{davis2005integrated,landwehr2005nfoil}
 \end{compactenum}
\\
\hline
\T \B \multirow{1}{*}{\rotatebox{0}{\normalsize \textbf{Section~\ref{sec:feature-selection}) Relational Feature Selection}}} 
\begin{compactenum}[\scriptsize$\bullet$\leftmargin=0em]
\item Correlation-based~\cite{guyon2003introduction}
\item Log-binning disagreement~\cite{milojevic2010power}
 \end{compactenum}
\\
\hline
\end{tabular*}
\end{center}
\vspace{-6mm}
\end{table}

\subsection{Relational Feature Classes}\label{sec:feature-classes}
We define the relational feature space with respect to the relational information used in the feature computation (i.e., edges/nodes only, graph+node/edge attributes, or non-relational information) giving rise to four main feature classes: structural features ($\mathcal{G}$), link-value features ($\mathcal{G}, \mX^{e}$), node-value features ($\mathcal{G}, \mX^{v}$), and non-relational features (which either uses $\mX^{e}$ or $\mX^{v}$ in the computation depending on if a non-relational link or node feature is being computed)\footnote{Clearly, these four classes of features can be recursively computed.}.
The information in parenthesis represents the information used in the feature computation.
It should be clear that both link features and node features can have features that are computed from each of the four classes of features.
Now, let us define more precisely these four feature construction classes for the links and nodes.

\subsubsection{Structural features} \label{sec:structural-features}
Structural features are computed using only the structure of the graph $\mathcal{G}$ (and not features).
Examples include traditional features like degree, clustering coefficient, betweenness, etc.
In general, any type of subgraph pattern (e.g., number of 2-star patterns, triangles,...) or path/walk-based measures may be used to generate a large number of features.
Let us note that these structural features can be computed for the links and nodes by a simple reinterpretation~\cite{transformingSRL:2012}.
Roles as defined mathematically in the past~\cite{lorrain1971structural,white1983graph,holland1981exponential} are based strictly on structural properties.
Hence, these types of structural features are the most important for capturing the traditional notion of roles~\cite{everett1994regular,hollandkathryn1983stochastic}.

\subsubsection{Link-value features} \label{sec:link-value-features}
Link-value features are computed using only the feature values of the links adjacent to the target node (or link)\footnote{The target node (or link) is the node for which the feature is being constructed.}.
This process may easily be generalized to be the links $\rho$-hops away (i.e., paths of length $\rho$) away.
Thus, the only information used in the computation is the graph $\mathcal{G}$ and the link features $\mX^e$.
As an example, given the feature-values of the adjacent links, one might apply some type of aggregate such as the mode to compute an additional feature. 
This feature class, like the others, can be used for computing node features or link features.
In addition, another more difficult to see link-value (or node-value) features can be constructed via a low-dimensional approximation of $\fX^e$ to generate additional features.

\subsubsection{Node-value features} \label{sec:node-value-features}
Similarly, node-value features are computed using only the feature values of the nodes that are adjacent (or a few hops away) from the target node.
Thus, a feature is considered a relational node-value feature if the feature values of nodes linked to the target node are used in the construction.
For instance, consider a political affiliation node attribute in Facebook, then given a node, we could construct a new node-value feature based on the mode of the values of the adjacent nodes.
More generally, we could use the nodes $k$-hops away to compute a new node-value feature.
As an aside, if $k>1$, then one may decay the weight of that node on the feature calculation with respect to their distance such that nodes that are further away from the target node are given less influence in the feature calculation.
Instead of the mode, one may choose to count the number of adjacent nodes of each political affiliation (e.g., number of liberal nodes $k$-hops away).

\subsubsection{Non-relational features} \label{sec:non-relational-features}
A non-relational feature is defined as a feature computed using only the non-relational features (i.e., attributes), ignoring any link-based information~\footnote{Let us note that links and nodes may have non-relational features.}.
Traditionally, the new feature value of that node is computed using the non-relational features of that node or the entire collection of nodes as done traditionally.
The key idea is that the link-based information is ignored.
Given a feature vector for an arbitrary node (or link), one might construct additional features by summing together that node's feature values, thresholding a single value, etc~\cite{kohavi1997wrappers}.
For instance, suppose there are two node features representing ``avg time online'' and ``avg number of Facebook posts'', then one may construct a new non-relational feature representing the sum of the two features.
If there is textual data, then topic models~\cite{steyvers2007probabilistic} may also be utilized to construct new non-relational features, e.g., representing the node's main topic.
The non-relational features can be used to better tune the role discovery process for a specific application.

\subsubsection{Discussion}
The feature-based roles can naturally incorporate any of the above types of features.
The only requirement is that for any set of link features, one must first use some type of feature operator (e.g., aggregate) to construct features on the node.
We discuss opportunities for discovering link roles in Section~\ref{sec:challenges}, that is assigning roles to individual links, instead of nodes.
However, any of the node features, including the non-relational features (i.e., attributes) can be used directly in the role computation by simply adding them to the node-feature matrix $\mX$.

\newcommand\JT{\rule{0pt}{2.2ex}}
\newcommand\JB{\rule[-1.2ex]{0pt}{0pt}}
\begin{table}
\vspace{-2mm}
\caption{
\vspace{-2mm}
\textbf{Relational Feature Operators for Roles}.
Features may be constructed from relational graph data using any of the following relational operators below.
As an example, subgraph pattern operators consist of both local graph features such as wedges or triangles and global features such as average shortest path between vertices.
} 
\vspace{-4mm}
\label{table:feature-ops}
\begin{center}
\begin{small}
\begin{tabular}{l !{\vrule width 0.5mm} p{48mm}}
\multicolumn{1}{l !{\vrule width 0.5mm}}{\large \textbf{Operators}} \JT \JB
& \textbf{\large Examples}
\\
\noalign{\hrule height 0.5mm}
\textbf{Rel. aggr.}~\cite{neville:srl00,getoorlinkbased2003} & 
\JT \JB\textsc{Mode}, \textsc{Mean}, \textsc{Count}, ...
\\ 
\hline

\textbf{Set ops}~\cite{aha:09} & 
\JT \JB Union, multiset, inters., ...
\\ 
\hline

\textbf{Subgraph pat.}~\cite{robins:sn07} & 
\JT \JB k-star, k-clique, k-motif, ...
\\ 
\hline

\textbf{Dim. redu.}~\cite{sarwar2000application,fodor2002survey} & 
\JT \JB SVD, PMF, NMF, ICA, PCA, ...
\\ 
\hline

\textbf{Similarity}~\cite{boriah30similarity,Lin98aninformation-theoretic} & 
\JT \JB Cosine sim, mutual info, ... 
\\ 
\hline

\textbf{Paths/walks}~\cite{niteshchawlakdd2010} &  
\JT \JB random-walks, k-walks, ...
\\ 
\hline

\textbf{Text analy.}~\cite{chang2009relational,rossi:10} & 
\JT \JB LDA, Link-LDA/PLSA, ... 
\\ 
\hline

\toprule
\end{tabular}
\end{small}
\end{center}
\vspace{-5mm}
\end{table}

\subsection{Relational Feature Operators for Roles}\label{sec:feature-ops}
At the heart of feature construction are the actual feature operators that will be used in the underlying search process.
These feature operators define the space of features that can potentially be searched over to construct a feature set, which in turn will ultimately be used to define the roles. 
The main classes of feature operators are categorized in Table~\ref{table:feature-ops}, along with examples of each.
Clearly, there is an infinite number of features that can be computed from these classes of feature operators.

Many of the feature operators can naturally be used to compute feature values for links and nodes.
In addition, the majority of feature operators can compute features using most of the previous classes of inputs from Section~\ref{sec:feature-classes}.
However, some of the operators that rely on non-relational information (i.e., text analysis), are obviously not applicable for constructing structural features, but can be applied for constructing link/node-value features and additional non-relational features (see Section~\ref{sec:feature-classes}).
In addition, some of these relational operators can be applied recursively (e.g, aggregates, set ops, clustering, dimensionality reduction) and in an iterative fashion (to compute additional recursive features).
For instance, we might compute the number of two-star patterns for a given node, then use the max relational aggregate operator, which would find the neighbor with the maximum number of two-star patterns and use this value as an additional feature-value.
Let us note that Section~\ref{sec:feature-classes} and~\ref{sec:feature-ops} defines the overall feature space while Section~\ref{sec:feature-search} defines how this space is explored.
See~\cite{transformingSRL:2012} for additional details.

\subsubsection{Discussion}
To construct meaningful, accurate and useful roles, it may be necessary to define a small subset of relational operators based on domain specific knowledge or assumptions. 
For instance, for social networks, operators that construct more local features (e.g., subgraph patterns) may be more useful whereas for technological networks other operators that construct global features may be more appropriate.
On the other hand, if the evaluation criteria for searching this space of features is tuned for a specific application or domain, then the feature search strategy should select the appropriate features (and providing a small subset of operators is not necessary).

\subsection{Feature Search Strategy} \label{sec:feature-search}
We previously defined the possible role feature space by specifying the raw feature inputs from Section~\ref{sec:feature-classes} (e.g., structural features, node/link-value, and non-relational features) and the relational operators to consider (Section~\ref{sec:feature-ops}).
The next step is to select an appropriate relational search strategy, which are usually either exhaustive, random, or guided search. 
An exhaustive strategy considers all features that are possible given the specified inputs and operators~\cite{neville:kdd03}, while
a random strategy will consider only a fraction of this space through some sampling strategy (See ~\cite{mcgovern2008spatiotemporal} for an example). 
A guided strategy uses a heuristic to identify the features that should be considered~\cite{davis2005integrated,landwehr2005nfoil}. 
In all three cases, each feature that is considered is subjected to some evaluation strategy that assesses its utility for representing roles (See Section~\ref{sec:feature-selection} for more details).
In Table~\ref{table:srl-systems}, we provide a summary of relational learning systems that may be refined for feature-based roles.
Since the majority of systems in Table~\ref{table:srl-systems} use guided search (instead of exhaustive or random), we list the name of the technique utilized.
Note the guided feature learning also includes methods that learn weights incrementally for the feature subspaces.
Usually the weights depend on the ability of learning novel and useful features from that subspace, e.g., all feature subspaces may be sampled uniformly in the first iteration, then biased in the subsequent iterations by the number of novel non-redundant features discovered.
Hence, the role feature learning is guided by the weight (or bias) assigned to each feature subspace and updated after every subsequent iteration.

As mentioned previously, roles are largely domain and application dependent, and thus the evaluation strategy that assesses features should usually be appropriately tuned.
Nevertheless, we can state some basic properties for constructing features that are suitable for computing generalized roles that match the previous definitions from Section~\ref{sec:foundations}.
Any arbitrary feature set constructed from a feature search technique should be: 

\medskip
\begin{compactenum}[$\triangleright$ \leftmargin=0em]
\item \textbf{Representative.} The set of features should be representative of the main structural patterns in the graph (or those of interest to a specific domain or application). 
\medskip \item \textbf{Minimal.} The set of features should be minimal. There should not be redundant features.
\end{compactenum}
\medskip

\noindent
However, as shown later in Section~\ref{sec:role-assignment}, the importance of these properties also depend on the technique used for assigning roles using the learned feature representation.
For instance, some low-dimensional approximation techniques should automatically handle redundant features as they would simply be represented together in a much lower-dimensionality, while the other non-representative features would be regarded as ``noise'' and essentially removed. 
The most important issue is that the main structural properties/patterns of the graph are captured by the constructed graph features.

One may construct features either manually or automatically.
Manual feature construction is essentially an exhaustive method that does not score or select features iteratively, and thus one needs to only select the types of features to construct (Section~\ref{sec:feature-classes}) and the feature operators (Section~\ref{sec:feature-ops}) to use over those types of features. 
In other words, the set of features are predefined based on expert knowledge of the problem, application and/or requirements.
One key advantage of manually choosing features is that the resulting roles will be easier to interpret since they are based on a set of finely tuned features.
Also, in cases where the problem/application is well-defined and there is plenty of domain knowledge, which is unlikely to change (thus not requiring strong generalization properties), then manual specification may result in stronger application-specific roles. 
The disadvantages of course are that an expert might miss a feature that is important to model or the known assumptions may change over time.
But more importantly, the time and monetary costs to actually perform this tuning may make it impractical for most tasks.

Alternatively, we may construct features automatically in a non-parametric fashion using a system~\cite{davis2005integrated,landwehr2005nfoil,neville:kdd03,rossi2012drc,mcgovern2008spatiotemporal}, which is the primary focus of this article.
For instance, Algorithm~\ref{alg:feature-construction-template} essentially searches a space of features automatically until no further novel features emerge. 
In that example, the feature space is defined by the set of primitive feature operators used initially (e.g., degree/egonet/k-core or other variants) and the set of recursive relational operators (e.g., sum, mode, etc.) selected.
At each iteration, new features are constructed, redundant ones are discarded, and this process repeats until there are no more novel/useful features being generated.
Other variants of the above are also possible.
This automatic approach is more appropriate for large-scale analysis where roles must generalize across networks and/or roles may not be well understood (limited assumptions).
The roles generated from the automatic feature construction are typically more difficult to interpret, but have the advantage of capturing arbitrary structural patterns.
The ability to capture novel structural patterns may be important for applications such as anomaly detection.
A more reasonable approach might be to have an expert manually tune a set of features and then use a system for automatically constructing additional features. 
This way, the roles are guaranteed to capture the properties warranted by the application (and expert) and also capture the main features of the graph data.

\subsection{Relational Feature Selection: Scoring \& Pruning}  \label{sec:feature-selection}
The previous section discussed strategies for searching over the space of relational features to generate a candidate set of features that capture the fundamental node-centric structural patterns (for assigning roles).
Now we must decide how to evaluate the role-based features, which consists of (i) scoring the candidate features and then (ii) selecting/pruning them using these scores and any additional knowledge.
The two fundamental goals of \textit{unsupervised feature selection} are reduction and denoising.
Reduction seeks to decorrelate and remove dependencies among features, whereas denoising attempts to find and reduce noisy features, therefore providing features that are more discriminative.

Let us start by defining the general notion of a similarity matrix\footnote{The similarity matrix may be viewed as a weighted similarity graph between the features.}.
Given a similarity function $\simfun$ and a set or matrix of features $\fX$, we define a similarity matrix as $\simmat = \forall (i,j) \in F \; , \simfun(\vx_i,\vx_j)$.
There are various similarity measures that can be used to evaluate role-based features such as Pearson correlation (and Spearman rank correlation)~\cite{cha2007comprehensive}, information gain~\cite{yao2003information}, gain ratio~\cite{aggarwal2012use}, gini-index~\cite{guyon2006feature}, and Kearns-Mansour criteria~\cite{kearns1998information}, among others~\cite{guyon2006feature,cha2007comprehensive,transformingSRL:2012}.
We also note that logarithmic binning is also useful for many sparse real-world networks with skewed node-level distributions (degree, number of triangles, etc) such as social and information networks~\cite{milojevic2010power}.
Other metrics that could be used include maximal information coefficient (MIC)~\cite{reshef2011detecting}, Mallows $C_{p}$~\cite{mallows2000some}, Bayesian information criterion (BIC)~\cite{hannan1979determination,schwarz1978estimating} and many others~\cite{shao1996bootstrap,george1993variable}.
In addition, Table~\ref{table:srl-systems} summarizes the feature evaluation used in a variety of relational learning systems found in the literature.

\newcommand\TE{\rule{0pt}{2.0ex}}
\newcommand\BE{\rule[-1.1ex]{0pt}{0pt}}

\begin{table}[t!]
\caption{
\vspace{-2mm}
\textbf{Systems for Searching and Selecting Features}
}
\vspace{-3mm}

\scriptsize
\begin{center}
\begin{tabulary}{1.0\linewidth}{RLl}
\toprule
\TT \BB 
\textbf{Proposed System}
& 
\textbf{Search method}
&  
\textbf{Feature evaluation}
\\
\midrule

\TE \BE \textbf{ \textbf{RPT}} ~\cite{neville:kdd03}
& Exhaustive 
& $\chi^2$ statistic/p-value
\\

\TE \BE \textbf{ \textbf{RDN-Boost.}}~\cite{natarajan2012gradient,khot2011learning}
& Exhaustive
& Weighted variance
\\

\TE \BE \textbf{ \textbf{ReFeX}}~\cite{refex2011kdd}
& Exhaustive
& Log-binning disagreem.
\\

\TE \BE \textbf{ \textbf{Spatiotemp. RPT}}~\cite{mcgovern2008spatiotemporal} 
& Random 
& $\chi^2$ statistic/p-value
\\

\TE \BE \textbf{ \textbf{SAYU}}~\cite{davis2005integrated}  
& Aleph
& AUC-PR
\\

\TE \BE \textbf{ \textbf{nFOIL}}~\cite{landwehr2005nfoil}
& FOIL 
& Conditional LL
\\

\TE \BE \textbf{ \textbf{SAYU-VISTA}}~\cite{davis:ijcai07}
& Aleph 
& AUC-PR
\\

\TE \BE \textbf{ \textbf{Discri. MLN}}~\cite{huynhDiscimStruct2008}
& Aleph++
& {\em m}-estimate
\\

\TE \BE \textbf{ \textbf{ProbFOIL}}~\cite{de2010probabilistic}
& FOIL
& {\em m}-estimate
\\

\TE \BE \textbf{ \textbf{kFOIL}}~\cite{landwehr2010fast}
& FOIL
& Kernel target-alignment
\\

\TE \BE \textbf{ \textbf{PRM stru. learn.}}~\cite{getoor:icml01}
& Greedy hill-clim.
& Bayesian model selection
\\

\TE \BE \textbf{ \textbf{TSDL}}~\cite{kok:05}  
& Beam search
& WPLL
\\

\TE \BE \textbf{ \textbf{BUSL}}~\cite{mihalkova2007bottom} 
& Template-based
& WPLL
\\

\TE \BE \textbf{ \textbf{PBN Learn \& Join}}~\cite{khosravi2010structure}
& Level-wise search
& Pseudo-likelihood
\\

\bottomrule
\end{tabulary}
\end{center}
\label{table:srl-systems}
\end{table}

Thus far, we have mostly presented unsupervised approaches that are based on the notion of a representative set of features that are non-redundant and minimal.
However, the feature learning process may be guided by the underlying assumptions and knowledge about the application-specific objective of the roles being constructed.
As an example, a candidate feature for roles may be evaluated by classification in an iterative fashion; if accuracy improves on a holdout set, then the feature may be added~\cite{davis2005integrated,davis:ijcai07}.
In other cases, features may be scored at each successive iteration, and then only the feature with the largest score may be retained~\cite{mihalkova2007bottom}.

Given the similarity matrix $\simmat$ containing similarity scores/weights between all such pairs of features, how should we decide on the relational features to select?
This problem may also be viewed as a relational role-based feature pruning problem.
For feature-based roles, the feature selection method is driven by the goal of obtaining a representative set of features that is also minimal (e.g., redundant and noisy features are removed).
This also has the additional benefit of improving the space-efficiency of the feature representation, which is important for large real-world networks.
The pruning is generally performed automatically using a threshold (e.g., 0.5), which defines the level at which two features are labeled as similar.
However, the pruning and similarity may be tuned for specific applications as well.

{\algrenewcommand{\alglinenumber}[1]{\scriptsize #1  }
\begin{figure}[t!]
\vspace{-2mm}
\begin{minipage}{1.0\linewidth}
\begin{algorithm}[H]
\caption{Role Feature Learning Template}
\label{alg:feature-construction-template}
{\footnotesize
\begin{algorithmic}[1]
\State {\bf Input:} \quad $G=(V,E,\X^{\rm attr})$ -- Initial graph and attributes, \label{algline:initial-data}
$\primops$ -- Set of primitive operators, \label{algline:prim-ops}
$\ops$ -- Set of relational iterative operators, \label{algline:ops}
$\sim(\cdot)$ -- Score function, 
$\maxiter$ -- Maximum number of iterations allowed, \label{algline:maxiter}
$\lambda$ -- Threshold for searching \label{algline:threshold}
\smallskip
\fontsize{8}{10}\selectfont
\State $F_{0} \leftarrow$ \textsc{Primitives}($G$,$\primops$)	\label{algline:primitive-features}
\State Let $\X$ be the feature data computed from the primitives \label{algline:initialize}

\For{$t \leftarrow 1$ {\bf to} \maxiter} \label{algline:prim-for-maxiter} 

	\State $F_{t},\X  \leftarrow ${\sc FeatureSearch}($F_{t-1}, \X, \ops)$ \label{algline:search} 
	\State $F_{t} \leftarrow F_{t} \cup F_{t-1}$ \label{algline:combine-features}

	\State $\mathcal{G}_F \leftarrow ${\sc CreateFeatureGraph}$(F_{t}, \X, \sim$, $\lambda$) \label{algline:feature-graph} 
	\State $\mathcal{C} \leftarrow$ Partition the feature graph $\mathcal{G}_F$ (e.g., conn. components) \label{algline:components}
	\For{ {\bf each} $\mathcal{C}_k \in |\mathcal{C}|$} \Comment{Prune features} 
		\State Find the earliest (or $\min$ corr.) feature $f_i$ s.t. $\forall f_j \in \mathcal{C}_k : i < j$.
		\State $F_{t} \leftarrow \big(F_{t} \setminus \mathcal{C}_k) \cup \{f_i\}$
	\EndFor	
	\State Remove features from $\X$ that were pruned (not in $F_{t}$) \label{algline:ensure-data}
	\If{$F_{t} = F_{t-1}$}  {\bf terminate search}\label{algline:converge} \Comment{no new features} 
	\EndIf
\EndFor
\State {\bf return} $\X$ and $F_t$ \Comment{feature matrix $\X \in \mathbb{R}^{n \times f}$ and  list $F_t$}
\algrule[0.3pt]

\Procedure{CreateFeatureGraph}{$F_{t-1}, \X, \sim)$, $\lambda$}
\State Set $\mathcal{G}_F = (V_F,E_F)$ -- the initial feature-graph
\State Set $V_F$ to be the set of features from $\fset$ and $E_F = \emptyset$

\For{{\bf each} pair of features $(f_i, f_j) \in F_{t-1}$}
	\If{$\sim(f_i, f_j) \geq \lambda$}
		Add edge $(i,j)$ to $E_F$
	\EndIf
\EndFor
\EndProcedure

\end{algorithmic}}
\end{algorithm}
\end{minipage}
\vspace{-3mm}
\end{figure}}

\subsection{Generalized Role Feature Learning Template} \label{sec:feature-construction-template}
Section~\ref{sec:feature-classes} and~\ref{sec:feature-ops} defined the space of relational features for roles whereas Section~\ref{sec:feature-search} and~\ref{sec:feature-selection} determined how this space is explored and evaluated.
The underlying decisions essentially determine the utility of the roles and therefore should be guided by knowledge and assumptions known prior about the specific-application in which the learned roles will be used.

We present in Algorithm~\ref{alg:feature-construction-template} a generalized (iterative) algorithm template for constructing a feature representation for role discovery from relational graph data.
Given a large graph and attributes, how do we learn a feature representation that captures the relational dependencies between attributes in the graph and the structural properties and patterns present in the graph data?
We propose an iterative graph feature discovery algorithm that learns a feature representation from a graph and any attributes. 
In particular, the learned features succinctly capture the main structural patterns/properties more compactly (non-redundant) while also revealing novel features.

The overall idea of Algorithm~\ref{alg:feature-construction-template} is to start from the relational data $G=(V,E,\X^{\rm attr})$ given as input, including graph data and attributes.
Using a set of primitive relational operators, the algorithm constructs new features that are added to the initial feature set $F_{0}$ (Line~\ref{algline:primitive-features}).
After the construction and pruning of the primitive features, the algorithm proceeds iteratively.
At each iteration, new features are constructed using a set of relational iterative operators $\ops$ (See Line~\ref{algline:search}).
The previous set of features $F_{t-1}$ are added to the new set of features $F_{t}$, thus $F_{t} \leftarrow F_{t} \cup F_{t-1}$ shown in Line~\ref{algline:combine-features}.
Now, we compute scores between all pairs of features and prune edges between features that are \textit{not} significantly correlated:
\[ E_{F} = \{ (f_i, f_j) \; | \; \forall (f_i, f_j) \in F \times F \text{ s.t. } \sim(f_i, f_j) > \lambda \} \]
This process results in a weighted feature graph where large edge weights represent dependencies between two features (Line~\ref{algline:feature-graph}).
Now, we use the weighted feature graph $\mathcal{G}_F$ to prune all redundant and noisy features from $F_{t}$.
This is done by first partitioning the feature graph (Line~\ref{algline:components}).
For partitioning we use connected components, though there are numerous other possibilities (e.g., largest clique).
Intuitively, each connected component represents a set of redundant features since edges represent dependencies.
Therefore, for each component, one may prune all but the earliest (or least correlated) feature.
Observe that the features learned at each iteration are guaranteed to be preserved in further subsequent iterations and therefore, $|F_1| \leq \cdots \leq |F_{t-1}| \leq |F_{t}|$.
At the end of each iteration, we ensure that the feature matrix $\X$ reflects the set of features $F_{t}$ (Line~\ref{algline:ensure-data}).
Hence, pruned features are removed from $\X$.
Finally, the iterative algorithm converges if no new features were constructed, hence $|F_{t}| = |F_{t-1}|$.
Otherwise, the algorithm proceeds iteratively, repeating the previous steps.
As an aside, Algorithm~\ref{alg:feature-construction-template} is well-suited for directed/undirected graphs, which may be weighted, timestamped, multi-typed, and contain an arbitrary number of node and edge attributes.
Many components of the algorithmic template are interchangeable and thus may be replaced/adapted for use with other techniques (e.g., backward feature selection, and others~\cite{guyon2006feature}).
Furthermore Algorithm~\ref{alg:feature-construction-template} is easily modified for extracting a list of previously learned features $F$ on another input graph (useful for transfer learning, etc).

\section{Role Assignment} \label{sec:role-assignment}
Section~\ref{sec:feature-construction} proposed a general role-based feature learning framework for automatically transforming the the graph data into a representative set of features that capture the fundamental notion of roles.
This section focuses on the second critical phase for discovering feature-based roles: how to assign nodes with similar feature vectors to the same role?
In particular, we survey two main categories of methods for assigning roles using the graph-based feature representation: role clustering methods (Section~\ref{sec:role-clustering}) or low-rank approximation techniques (Section~\ref{sec:low-rank-approx}).
Techniques to automatically select the best number of roles are discussed in Section~\ref{sec:model-selection}.
For each category of methods, there are also both soft role assignment and hard role assignment techniques (Section~\ref{sec:hard-vs-soft}).
We also note that some role assignment methods learn role definitions and thus generalize in the sense that roles learned on one network may be extracted on another arbitrary network (See Table~\ref{table:applications} for potential applications).
Other useful properties are discussed below whenever appropriate.

\subsection{Role Clustering} \label{sec:role-clustering}
There are many clustering algorithms that can be used to assign nodes to their corresponding roles using the graph features $\mX$.
The two primary types are hierarchical clustering algorithms (e.g., agglomerative
or divisive clustering)~\cite{murtagh2012algorithms} and partitioning algorithms 
such as k-means, k-medoids~\cite{berkhin2006survey,zhu2006semi},
and self-organizing maps~\cite{kohonen1990self}.
Many of the traditional clustering methods such as k-means are hard-clustering techniques.
This is in contrast to soft-clustering methods which allow nodes to be in multiple clusters.
A few classical methods are fuzzy C-means~\cite{bezdek1984fcm} or types of Gaussian Mixture Models~\cite{rasmussen2000infinite}, among others~\cite{erosheva2005bayesian}

\subsection{Low-rank Approximation}
\label{sec:low-rank-approx}
The other way to compute roles from a large feature matrix $\mX \in \mathbb{R}^{ n \times f}$ is to select a relatively small number $r$ where $r \leq f$  (usually automatically, see \ref{sec:model-selection}) and compute a low rank-$r$ matrix $\hat{\mX}_r$ that best approximates the original feature matrix with respect to any standard matrix norm.
There are many dimensionality reduction methods that can be used for this purpose.
A few of the most common methods are SVD~\cite{golub1970singular}, Principal Component Analysis (PCA)~\cite{abdi2010principal}, Kernel-PCA~\cite{scholkopf1997kernel}, MDS~\cite{kruskal1964nonmetric}, spectral decomposition, NMF~\cite{wang2013nonnegative,lee1999learning}, CUR~\cite{mahoney2009cur}, Factor Analysis, Probabilistic Matrix Factorization (PMF)~\cite{salakhutdinov2008probabilistic}, Independent Component Analysis (ICA)~\cite{du2014independent,comon1994independent}, among many others.
The majority of low-rank approximation techniques have a matrix $\mU$ in SVD (and $\mW$ in NMF) where each row represents the role-memberships for each node. 
While not all of the clustering algorithms allow nodes to be in multiple roles, nearly all the low-rank approximation techniques can be thought of as assigning nodes a role-membership.
We discuss and provide two examples below.

\textit{Singular-value Decomposition (SVD).} 
Let $\mX \in \mathbb{R}^{ n \times f}$ be the node by feature matrix, then we decompose $\mX$ into three matrices using the SVD, 
$\mX = \mU \mS \mV^T$ where $\mU \in \mathbb{R}^{ n \times f}$, $\mS \in \mathbb{R}^{ f \times f}$, $\mU \in \mathbb{R}^{ f \times f}$.
The matrix $\mS$ contains the singular values located in the 
$(i,i)_{1,..,f}$ cells in decreasing order of magnitude and all other cells contain zero. 
The eigenvectors of $\mX\mX^T$ make up the columns of $\mU$ and the eigenvectors of $\mX^T \mX$ make 
up the columns of $\mV$. 
The matrices $\mU$ and $\mV$ are orthogonal, unitary and span vector spaces of dimension n and f, respectively. 
The columns of $\mU$ are the  principal directions of the features and the rows of $\mV^T$ are the principal directions of the nodes. 
The principal directions are ordered according to the singular values and therefore according to the importance of their contribution to $\mX$.

The SVD is used by setting the insignificant $f-r$ singular values to zero, which implies that we approximate the matrix $\mX$ by a matrix $\mX_r = \mU_r \mS_r \mV_r^T$.
A fundamental theorem by Eckart and Young~\cite{eckart1936approximation} states 
that $\mX_r$ is the closest rank-$r$ least squares approximation of $\mX$. 
This theorem allows us to set the insignificant singular values to zero and keep only the few influential singular values.
The columns of $\mU_r \in \mathbb{R}^{n \times r}$ represent the most significant roles while each row of $\mU_k$ represents a node's membership in each of the roles.
The singular values in $\mS$ provide contribution scores for each of the roles in $\mU_r$ and $\mV_r^T$. 
The columns of $V_r$ represent how roles contribute to the features.
Essentially, the underlying roles were hidden in the large set of features, but when we reduce the dimensionality, the latent roles become apparent as similar graph features are combined.
There are also many other methods similar to SVD that could be easier to interpret~\cite{mahoney2009cur}.

\textit{Non-negative Matrix Factorization (NMF).} 
We also provide an example of computing roles using NMF. 
Given a node-feature matrix, we generate a rank-r approximation $\mW \mH \approx \mX$ where each row of $\mW \in \mathbb{R}^{n \times r}$ represents a node's membership in each role and each column of $\mH \in \mathbb{R}^{r\times f}$ represents how membership of a specific role contributes to estimated feature values. 
More formally, given a nonnegative matrix $\mathbf{\mX} \in \mathbb{R}^{n \times f}$ and a positive integer $r < \min(n,f)$, find nonnegative matrices $\mathbf{\mW} \in \mathbb{R}^{n \times r}$ and $\mathbf{\mH} \in \mathbb{R}^{r \times f}$ that minimizes the functional,
$f(\mathbf{\mW}, \mathbf{\mH}) = \frac{1}{2}||\mathbf{\mX} - \mathbf{\mW}\mathbf{\mH}||^{2}$
Let us note that NMF is often ``easier'' to interpret than SVD, due to the non-negativity constraint.
As an aside, role clustering methods and low-rank approximation methods may sometimes overlap, e.g., there has been a lot of work showing the equivalence between NMF and k-means clustering~\cite{ding2005equivalence,ding2008equivalence}.

\textit{Discussion:} 
Depending upon the application, expected data characteristics, and computational requirements (i.e., memory, runtime, or scalability constraints), the role-based low-rank approximation methods in the framework are flexible for the following:
(i) similarity/objective function (e.g., Frobenius norm, KL-divergence),
(ii) regularization terms if warranted (e.g., sparsity constraints, L2, etc), and
(iii) solver (e.g., Multiplicative update, SGD, ALS).
For instance, suppose roles are assigned via an NMF-based approach using KL-divergence with L2 regularization terms and a CCD-based solver.
One may also add sparsity and other constraints~\cite{heiler2006controlling,cichocki2008flexible,cichocki2007novel} to these approaches to better adapt the roles for specific applications~\cite{gilpinguided}.
Furthermore depending on the input data and application, many techniques exist for improving the semantics of roles and their interpretation (e.g., pre/postprocessing techniques may help avoid/interpret negative coordinates).
We also note that many of these techniques may also be used for learning roles over a time series of graphs, see~\cite{rossi2013dbmm-wsdm}.
For dynamic networks, one may also represent the time series of graphs as a tensor where roles can now be learned using a tensor factorization method~\cite{kim2007nonnegative,de2000best,de2009survey,cichocki2008advances,yilmaz2010probabilistic}.

\subsection{Model Selection: Choosing the number of roles} \label{sec:model-selection}
Many techniques have been proposed for selecting the appropriate number of roles.
Some of these techniques are heuristic while others have a more fundamental basis in statistics (e.g., AIC~\cite{akaike1974new}) and information theory, e.g., Minimum Description Length (MDL)~\cite{cook2007graph,grunwald2007minimum,rissanen1978modeling}.
In this section, we survey a few of these approaches and discuss details of each and how they could be used in the context of role discovery.

\vspace{1mm} \noindent \textit{Selecting number of clusters.}
There has been a substantial amount of research on selecting the number of clusters automatically~\cite{tibshirani2001estimating,pelleg2000x,dudoit2002prediction,celeux1996entropy,sugar2003finding}.
The classical clustering methods such as hierarchical clustering~\cite{salvador2004determining} and k-means~\cite{pelleg2000x} have been adapted to select the number of roles automatically (using various criteria).

\vspace{1mm} \noindent \textit{Selecting the number of dimensions for low-rank approximation.}
Many methods have been developed for this purpose~\cite{dray2008number,eastment1982cross,jackson1993stopping,wold1978cross}.
For example, using SVD or other related techniques, it is common in information retrieval to select 50 or 100 eigenvectors as their importance typically follows a geometric distribution (in most types of data)~\cite{manning2008introduction}.
In general, the number of significant eigenvalues from the feature matrix will be much less for feature-based roles.
Empirically, the number of roles found in practice is usually quite small, between 2 and 15 roles are usually found for a variety of network types~\cite{dbmm11-tr}.

{\algrenewcommand{\alglinenumber}[1]{\scriptsize #1  }
\begin{figure}[t!]
\vspace{-2mm}
\begin{minipage}{\linewidth}
\begin{algorithm}[H]
\caption{Role Model Selection} 
\label{alg:role-model-selection}
{\footnotesize
\begin{algorithmic}[1]
\State Set mincost = $\infty$, failed = 0, trials = $\tau$
\State Set $\mW_0$ and $\H_0$ to be random matrices (normal dist.)
\State Scale  $\mW_0$ and $\H_0$ by $\max$ value in $\X$
\For{$r = 1$ {\bf to} $\min(n,f)$}
	\State Learn model $D(\X | \mW_0[:,r], \H_0[r,:])$ \label{algline:role-model-selection-slice}
	\State Compute $\mathrm{cost}$ of model via criterion (e.g., MDL, AIC)
	\If{cost $<$ mincost} \Comment{model improves likelihood} 
		\State Set mincost = cost, reset failed to 0, and save model	
	\Else \quad set failed = failed + 1
	\EndIf
	\If{failed $\geq$ trials}
		\State terminate search and return model
	\EndIf
\EndFor
\end{algorithmic}}
\end{algorithm}
\end{minipage}
\vspace{-4mm}
\end{figure}}

We provide a general role model selection template in Algorithm~\ref{alg:role-model-selection} for automatically selecting the number of roles for low-rank approximation techniques.
Intuitively, Algorithm~\ref{alg:role-model-selection} greedily increases the number of roles as long as the cost of the role model decreases (or likelihood improves using that number of roles).
Note that $\mW_0$ and $\H_0$ are the initial randomized matrices whereas $\mW_0[:,r]$ and $\H_0[r,:]$ in line~\ref{algline:role-model-selection-slice} are the first $r$ columns and rows from the initial $\mW_0$ and $\H_0$, respectively.
One may also use MDL to automatically determine the number of structural roles.
Intuitively, learning more roles increases model complexity but decreases the error (and vice versa). 
In this way, MDL selects the number of roles $r$ (representing structural patterns) such that the model complexity (number of bits) and model errors are balanced. 
Note the number of bits is typically $\log_{2}(\mathsf{number \; of \; parameters})$.
Naturally, the best model minimizes, $number \; of \; bits + errors$. 
Note that $\tau=5$ is a heuristic used in some optimization schemes to represent a finite number of forward greedy steps~\cite{jalali2011learning}. This approach provides statistical guarantees for the estimated model.

Importantly, Algorithm~\ref{alg:role-model-selection} may also serve as a basis for a more guided or supervised approach to selecting the number of roles (i.e., driven by an application-specific objective function).
For instance, if we were interested in classification, then one may learn the best model that maximizes the cross-validated AUC.

\subsection{Hard and Soft Role Assignments} \label{sec:hard-vs-soft}
\begin{figure}[h!]
\centering
\hspace*{-5mm}\subfigure[Hard Assignment]{
	\includegraphics[width=2.2in]{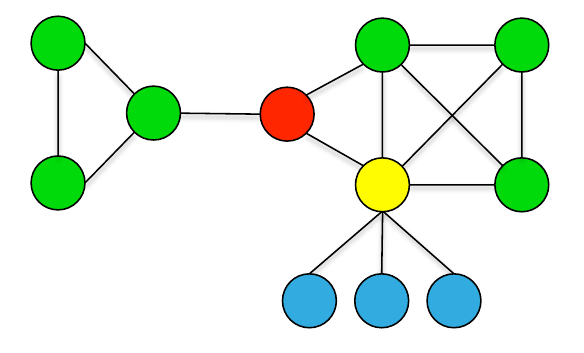}
	\label{fig:hard-assignment}
}
\hspace*{-5mm}\subfigure[Soft Assignment]{
	\includegraphics[width=2.2in]{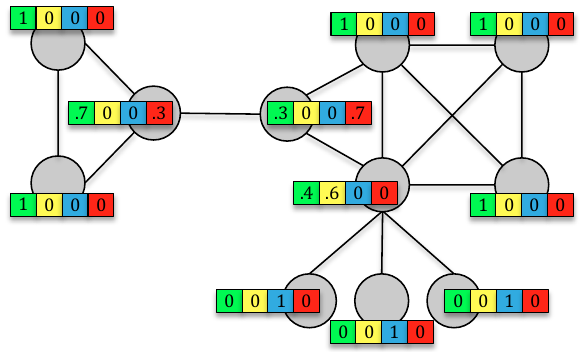}
	\label{fig:soft-assignment}} 
\caption{
\textbf{Example of Hard and Soft Role Assignments}.
Fig.~\ref{fig:hard-assignment} demonstrates hard role assignments to the nodes.
In particular, nodes are assigned the roles of 
\textcolor{black}{clique}, 
\textcolor{black}{bridge}, 
\textcolor{black}{star-center}, and 
\textcolor{black}{peripheral} role.
In Fig~\ref{fig:soft-assignment}, each node is assigned a distribution of roles s.t. $\sum r_i = 1$.
In some cases, this may be beneficial.
For example, the vertex in Fig.~\ref{fig:hard-assignment} that is assigned the bridge role also participates in a clique of size three.
In another example, the vertex in Fig.~\ref{fig:hard-assignment} that is assigned the star-center role is also participating in a clique of size four.
In both these cases, a hard assignment may not make sense, instead a distribution of roles should be assigned to accurately represent the node roles.
}
\vspace{-2mm}
\label{fig:hard-vs-soft-assignment}
\end{figure}

Hard role assignment refers to a vertex (or edge) being assigned to only a single role~\cite{nowicki:01,nowicki2001estimation,batagelj2004generalized}, whereas soft role assignment allows for vertices to play multiple roles (mixed-membership)~\cite{blei:jmlr08,rossi2012role-www,fu2009dynamic,xing2010state}.
See Figure~\ref{fig:hard-vs-soft-assignment} for an illustration.
Depending on the situation and constraints, one may make more sense than the other.
For instance, soft role assignment is typically more expensive in terms of storage requirements.
For a dynamic graph where each snapshot graph $G_t$ represents 1 minute (timescale), then it may be impractical to allow a vertex to play multiple roles.
As an example, if the vertices represent individuals, then it is unlikely that any individual would be playing more than a single role, since the time scale is so small.
However, if the snapshot graph $G_t$ represents 1 day of activity, then individuals are likely to play multiple roles.
As an example, an individual is likely to spend a good portion of the day in the ``work role'', then after work they may transition into the wife or mother role.

\begin{compactitem}[{\footnotesize $\bullet$}]
\item \textbf{Soft Role Assignment Methods}. 
Matrix factorization techniques allow for soft role assignments.
These include the typical methods of SVD, NMF, PMF, or any other factorization.
Gaussian mixture models also assign soft roles to vertices (or edges).

\item \textbf{Hard Role Assignment Methods}. 
The classical k-means and the variants are hard role assignment methods as these typically assign a data point to a single centroid (role).
\end{compactitem}

\section{Discussion and Challenges}\label{sec:challenges} 
This section discusses additional issues in role discovery and highlights important directions and challenges for future work.
A few of the tasks for which roles may be helpful are summarized in Table~\ref{table:applications}.

\subsection{Dynamic and streaming graphs}\label{sec:dynamic-roles}
Most role discovery methods introduced in this article are for static networks.
However, networks are not static, but are naturally dynamic and continuously streaming over time and thus role discovery methods for dynamic and streaming networks are of fundamental practical and theoretical importance.
Modeling ``dynamic'' roles are important for many practical applications (e.g., recommendation) due to the fact that user preferences and behavior are fundamentally tied to time.
Intuitively, methods for learning roles in dynamic networks must leverage the notion that recent user preferences are more predictive than preferences in the distant past.

There have only been a few approaches for dynamic feature-based roles.
More specifically, \textsc{Role-Dynamics} learns features from the time series of graph data, then assigns roles using those features.
Using the learned feature and role definitions, they now extract feature-based roles in a streaming fashion for detecting graph-based anomalies~\cite{rossi2012role-www}.
This approach uses simple local primitives (i.e., degree/egonet-based features) with  $\{mode, sum\}$ operators and used \textsc{nmf+mdl} for assigning roles.
More recently, \textsc{dbmm} extended that approach by modeling the global network role transitions as well as learning local role transition models for each node~\cite{rossi2013dbmm-wsdm}.
There remains many challenges and opportunities for dynamic feature-based roles.
For instance, one may instead represent the sequence of node-feature matrices $\mX_1, \mX_2, ..., \mX_t$ as a tensor and use tensor factorization techniques to learn the role definitions\cite{kim2007nonnegative,de2000best,de2009survey,cichocki2008advances,yilmaz2010probabilistic}.
Moreover, we may also utilize non-negative tensor factorization (NTF) techniques~\cite{kim2007nonnegative,friedlander2008computing} for grouping the features and thus providing a time-series of role mixed-memberships.
There are also regularized non-negative tensor factorization approaches~\cite{cichocki2007regularized} and constraint-based approaches (including sparsity and other constraints)~\cite{heiler2006controlling,cichocki2008flexible,cichocki2007novel} that can be used to better adapt the roles for specific applications.

In terms of graph-based roles, Fu~\etal proposed an approach based on MMSB for dynamic networks called dMMSB~\cite{fu2009dynamic, xing2010state} that essentially characterizes how the roles in MMSB change over time.
However, dMMSB does not scale to large networks found in the real-world~\cite{xing2010state} (See Section~\ref{sec:scaling-up}).
The other disadvantage is in the assumption of a specific parametric form, whereas feature-based roles are more flexible in the types of roles expressed and that roles are data-driven/non-parametric and thus can easily adapt based on stream/dynamic characteristics.

Online role discovery methods that assign and update role memberships as new data arrives remains an open area of research~\cite{vijayakumar2005incremental}.
These methods are not only required to update role memberships, but must incrementally update the underlying role model and definitions.
In feature-based roles, the set of features may also become stale and new representative features may need to be learned in an incremental fashion.
Additional work may investigate techniques to automatically learn the appropriate time scale, lag of the time series to use, and parameters for exponential smoothing.

\subsection{Evaluation and Guidance}
Roles have traditionally been used as a descriptive modeling tool to understand the ``roles''  played by actors in the network and has been limited to relatively small networks.
This makes it difficult to understand and evaluate the benefits of the various methods.
The previous work discovers roles, then uses them for applications (e.g., link prediction, anomaly detection).
However, there has yet to be any work that ``guides'' the roles \textit{during learning} by the end goal or application.
Ideally, the learned roles should be guided by the final goal and should adapt based on the application-specific goal.
For instance, one might adapt an approach that discovers roles that maximize classification accuracy on a hold-out data set.
In that case, classification accuracy is used as a surrogate evaluation measure that is carried out during the role discovery process (e.g., a role and its features are retained if classification accuracy increases).
These types of approaches are not only important for learning more goal-oriented roles, but may serve as a fundamental basis for evaluating role discovery methods.

There has yet to be any attempt to develop an evaluation framework with the aim at understanding the algorithmic tradeoffs including accuracy (e.g., approx. error, evaluation measure, interpretation, etc) and efficiency.
Accuracy and efficiency need to be carefully balanced by the user depending on the restrictions/knowledge of the domain and the end goal (application) of role discovery.

\subsection{Scaling up role discovery methods}\label{sec:scaling-up}
The majority of traditional graph-based role methods were only suitable for relatively small networks~\cite{fu2009dynamic, xing2010state}.
Recently, there has been work on scaling up generalized block models~\cite{chan2011increasing} as well as stochastic blockmodel variants using downsampling and a fast stochastic natural gradient
ascent algorithm for variational inference~\cite{yin2013scalable,ranganath2013adaptive}.
Though, there still remains a lot of work to be done in this area, for instance, scaling up the methods further for graphs of billions of edges, adapting other methods using these fast inference procedures, and evaluating and comparing the accuracy of the approximation methods and their scalability in terms of parallel speedup.

Despite the fact that feature-based role methods are in general much more efficient and scalable than graph-based roles, there has yet to be any parallel feature-based role methods.
Though, a systematic investigation into these methods and relative parallel speedups would be extremely useful.
We note that role features in Algorithm~\ref{alg:feature-construction-template} may be computed independently for each node in parallel while role definitions may also be learned in parallel~\cite{lopes2010non,yu2012scalable}.
Downsampling (or network sampling) is another promising direction for feature-based role methods~\cite{nk2012network}, for instance, feature definitions may be learned on a much smaller sampled network, then roles may be assigned to the sampled nodes based on this feature representation.

Additionally, methods for learning a set of space-efficient features remains a challenge.
In some cases, the features learned in past work required more memory to store than the sparse graph itself and thus impractical for massive graphs.
A promising direction would be to design methods for learning sparse features or better pruning methods to reduce this cost.
Further, there has yet to be any methods for learning (graph or feature-based) roles in a distributed setting (using MPI/MapReduce) as well as role discovery methods suitable for GPU parallelization (i.e., massively multicore architectures with limited memory), yet both are critical for practical settings in the age of big data~\cite{trelles2011big}.

\subsection{Additional Directions}
Recently, there has been some work on incorporating graph and textual data into roles~\cite{mccallumrole:07,danilevsky2013entity}.
Additional work is needed to systematically understand the impact and utility of such approaches.
Similarly, no one has included the latent topics as attributes in a relational feature learning system to generate additional novel features to improve the learned roles.

Incorporating initial attributes into the feature learning system (instead of only the graph) for learning roles is another promising area of research.
It is widely known that feature engineering is the most important factor contributing to the success of machine learning algorithms in practice~\cite{domingos2012few}.
Naturally, the same is true for feature-based roles and therefore using the attributes (or external interest) as a means to guide the feature learning system provides a unique opportunity to mesh the graph structure with external attributes allowing for a more guided approach to role learning.

There has not been any work on role discovery methods for heterogeneous networks.
Nevertheless, many of the methods introduced in this article may be adapted for multiple node or edge types, though there has yet to be any investigation or evaluation of such methods for practical applications.

Feature learning systems for graph data depend heavily on the ``correctness'' of the observed links~\cite{transformingSRL:2012}.
In many cases, the links in the graph may be difficult to observe, noisy, and may be ambiguous to the end goal of role discovery~\cite{caceres2014boosting}.
Thus, preprocessing the graph data (e.g., weighting the links) may substantially improve features learning via the role-based feature learning systems.
In addition, previous work has regarded all relational feature operators uniformly, though in practice some operators may be more useful than others.
Weighting the features via the relational operators that constructed them is likely to be useful for fine-tuning role discovery for certain applications and has yet to be explored.

Feature-based roles also has connections to the increasing important problem of deep learning~\cite{marc2007sparse,bengio2009learning}.
Though most work has focused on image data, we may view the feature-based roles as a generalized graph-based deep learning approach.
In particular, this article introduced a framework for learning a two-layered representation based on the graph structure.
Recently, additional layers were learned on top of the feature-based roles to model the ``role dynamics''~\cite{rossi2013dbmm-wsdm}.
In particular, they learned a global layer representing how the roles change over time as well as local multi-level layers for each node representing how the roles of a node change over time.

Semi-supervised role discovery methods is another promising direction that has yet to be addressed.
For instance, one may compute strict properties of the nodes in the graph  and use these to label examples of roles in the graph for which other roles can be learned and extracted.
These could now be used in a semi-supervised fashion to help moderate the role discovery algorithm leading to roles that are more interpretable and useful.

Finally, the problem of computing link roles has yet to be addressed.
However, many of the previous techniques for feature-based roles may be adapted for computing link-based roles.

\ifCLASSOPTIONcompsoc
  \section*{Acknowledgments}
\else
  \section*{Acknowledgment}
\fi
The authors would like to thank the reviewers for helpful comments.
This work is supported by the NSF GRFP Fellowship under Grant No. DGE-1333468.
\ifCLASSOPTIONcaptionsoff
  \newpage
\fi

\bibliographystyle{IEEEtran}
\bibliography{roles-tkde14-final}


\end{document}